\documentclass[a4paper,11pt,amsmath,amssymb,floatfix,nofootinbib]{article}
\usepackage{jheppub} 

\pdfoutput=1 

\usepackage[T1]{fontenc} 
\usepackage{color}   
\usepackage{graphicx,float}
\usepackage{dcolumn}
\usepackage{bm}
\usepackage{slashed}
\usepackage{float}

\begin{document}

\def\bb    #1{\hbox{\boldmath${#1}$}}
 \def\oo    #1{{#1}_0 \!\!\!\!\!{}^{{}^{\circ}}~}  
 \def\op    #1{{#1}_0 \!\!\!\!\!{}^{{}^{{}^{\circ}}}~}
\def\blambda{{\hbox{\boldmath $\lambda$}}} 
\def\eeta{{\hbox{\boldmath $\eta$}}}
\def\bxi{{\hbox{\boldmath $\xi$}}} 
\def\bzeta{{\hbox{\boldmath $\zeta$}}}
\def\sD{D \!\!\!\!/}
\def\sd{\partial \!\!\!\!/}
\def\qcd{{{}^{\rm QCD}}}   
\def\qed{{{}^{\rm QED}}}   
\def\2d{{{}_{\rm 2D}}}         
\def\4d{{{}_{\rm 4D}}}         





\title{\boldmath Open string  QED meson description of the X17   particle
 and dark matter}

\author{Cheuk-Yin Wong} 
\affiliation{Physics Division, Oak Ridge National Laboratory,
Oak Ridge, Tennessee 37831, USA}

\emailAdd{wongc@ornl.gov}


%
%
%


\begin{abstract}
{
As a quark and an antiquark cannot be isolated, the intrinsic motion
of a composite $q \bar q$ system in its lowest-energy states lies
predominantly in 1+1 dimensions, as in an open string with the quark
and the antiquark at its two ends.  Accordingly, we study the
lowest-energy states of an open string $q\bar q$ system in QCD and QED
in 1+1 dimensions.  We show that $\pi^0, \eta$, and $\eta'$ can be
adequately described as open string $q\bar q$ QCD mesons.  By
extrapolating into the $q\bar q$ QED sector in which a quark and an
antiquark interact with the QED interaction, we find an open string
isoscalar $I(J^\pi)$=$0(0^-)$ QED meson state at 17.9$\pm$1.5 MeV and
an isovector $(I(J^\pi)$=$1(0^-), I_3$=0) QED meson state at
36.4$\pm$3.8 MeV.  The predicted masses of the isoscalar and isovector
QED mesons are close to the masses of the hypothetical X17 and E38
particles observed recently, making them good candidates for these
particles.  The decay products of QED mesons may show up as excess
$e^+e^-$ and $\gamma \gamma$ pairs in the anomalous soft photon
phenomenon associated with hadron productions in high-energy
hadron-proton collisions and $e^+$-$e^-$ annihilations.  Measurements
of the invariant masses of excess $e^+e^-$ and $\gamma \gamma$ pairs
will provide tests for the existence of the open string $q\bar q$ QED
mesons.  An assembly of gravitating QED mesons are expected to emit
electron-positron pairs and/or gamma rays and their decay energies and
lifetimes will be modified by their gravitational binding energies.
Consequently, a self-gravitating isoscalar QED meson assembly whose
mass $M$ and radius $R$ satisfy $(M/M_\odot)/(R/R_\odot) \gtrsim 4.71
\times 10^5$ will not produce electron-positron pairs nor gamma rays
and may be a good candidate for the primordial dark matter.
}
\end{abstract}

\maketitle \flushbottom

\setcounter{footnote}{0} \tableofcontents

\section{Introduction}

Recent observation \cite{Kra16} of a light, neutral boson decaying
into an $e^+e^-$ pair with a mass of about 17 MeV, in the decay of the
$I(J^\pi)$=$0(1^+)$ state of $^8$Be at Atomki, has generated a great deal of
interest
\cite{Zha17,Fen16,Fro17,Bat15,Ros17,Ell16,Alv18,Mun18,Ban18,Pad18}.
Supporting evidence for this hypothetical X17 particle has been
reported recently in the decay of the excited $I(J^\pi)$=0($0^-)$
state of $^4$He \cite{Kra19}.  Earlier observations of similar
$e^+$$e^-$ pairs with invariant masses between 3 to 20 MeV in the
collision of nuclei with emulsion detectors have been reported
\cite{El88,El96,deB96,deB05,Jai07,deB10}.  Different theoretical
interpretations, astrophysical implications, and experimental searches
have been presented
\cite{Zha17,Fen16,Fro17,Bat15,Ros17,Ell16,Alv18,Mun18,Ban18,Pad18}.
However, a definitive description of the X17 particle has not yet
emerged.

The observations of the $e^+e^-$ pair with an invariant mass of about
17 MeV \cite{Kra16,Kra19} may appear perplexing, equally perplexing
have been numerous observations of excess $e^+$$e^-$ pairs, labeled as
``anomalous soft photons'', whenever hadrons are produced in
high-energy $K^+ p$ \cite{Chl84,Bot91}, $\pi^+ p$ \cite{Bot91}, $\pi^-
p$ \cite{Ban93,Bel97}, $pp$ collisions \cite{Bel02}, and $e^+$$e^-$
annihilations \cite{DEL06,DEL08,Per09,DEL10}.  Specifically in the
DELPHI exclusive measurements in the decay of $Z^0$ in $e^+ e^-$
annihilations, the excess $e^+$$e^-$ pairs have been observed to be
proportionally produced when hadrons (mostly mesons) are produced
\cite{Per09,DEL10}, and they are not produced when hadrons are not
produced in high-energy $e^+$+ $e^-$$\to$ $\mu^+$+ $\mu^-$
bremsstrahlung \cite{DEL08}.  The transverse momenta of the excess
$e^+$$e^-$ pairs lie in the range of a few MeV/c to many tens of
MeV/c, corresponding to a mass scale of the anomalous soft photons in
the range from a few MeV to many tens of MeV.

It happens as if the X17 particle and
the anomalous soft photons are not perplexing enough, there occurs in addition
the perplexing E38 boson particle with a
mass of about 38 MeV observed in the $\gamma \gamma$ invariant mass
spectrum in high-energy $p$C, dC, dCu reactions at Dubna \cite{Abr12,Abr19}.
The extra-ordinariness of these objects at the mass scale of many tens of
MeV appears to place them outside the domain of the
Standard Model.  It is nonetheless important to explore here whether
there may be a Standard Model description that can link these three perplexing objects
in a coherent framework.

Many different models have been presented to describe the anomalous
soft photons as arising from quantized bosons \cite{Won10,Won11,Won14}
or from a continuous spectrum \cite{Van89}-\cite{Kha14}.  We shall
focus our attention on the quantized boson description of
\cite{Won10,Won11,Won14} which has the prospect of linking the
anomalous soft photons with the X17 and E38 particles.  We note that
owing to the simultaneous and correlated production alongside with
hadrons, a parent particle of the anomalous soft photons is likely to
contain some elements of the hadron sector, such as a light quark and
a light antiquark\footnote{ The elements of the hadron sector comprise
  of $u$, $d$, $c$, $s$, $b$, $t$ quarks, antiquarks, and gluons.  The
  mass scale of the anomalous soft photons excludes all but the $u$
  and $d$ quarks and antiquarks as possible constituents of the parent
  particles of the anomalous soft photons.}.  The quark and antiquark
carry color and electric charges and they interact mutually with the
quantum chomodynamical (QCD) and quantum electrodynamical (QED)
interactions.  A parent particle of the anomalous soft photons cannot
arise from the quark-antiquark pair interacting with the QCD
interaction, because such an interaction will endow the pair with a
mass much greater than the mass scale of the anomalous soft photons.
We are left with the possibility of the quark and the antiquark
interacting with the QED interaction.  Such a possibility is further
reinforced by the special nature of a confining gauge interaction, for
which the greater the strength of the attractive confining
interaction, the greater will be the mass of the composite particle it
generates (see Eq.\ (\ref{eq1}) below), in contrast to a non-confining
interaction in which the effect is just the opposite.  Relative to the
QCD interaction, the QED interaction will bring the quantized mass of
a $q\bar q$ pair to the lower mass range of the anomalous soft
photons.  It was therefore proposed in \cite{Won10} that a quark and
an antiquark in a $q\bar q$ system interacting with the QED
interaction may lead to new open string bound states (QED-meson
states) with a mass of many tens of MeV. These QED mesons may be
produced simultaneously with the QCD mesons in the string
fragmentation process in high-energy collisions
\cite{Chl84,Bot91,Ban93,Bel97,Bel02,DEL06,DEL08,Per09,DEL10}, and the
excess $e^+e^-$ pairs may arise from the decays of these QED mesons.
The predicted mass of the isoscalar $I(J^\pi)=0(0^-)$ QED meson is
close to the X17 mass of about 17 MeV.  It is natural to inquire
whether the hypothetical X17 particle may be the isoscalar 0(0$^-$)
QED mesons predicted in \cite{Won10}.  It is also useful to inquire
whether there can be additional experimental tests to confirm or
refute such proposed QED mesons.  For example, the mass of the
isovector $(I(J^\pi)=0(0^-),I_3=0)$ QED meson predicted in
\cite{Won10} is close to the mass of the hypothetical E38 boson
observed with an invariant mass of about 38 MeV in high-energy $p$C,
$d$C, $d$Cu reactions at Dubna \cite{Abr12,Abr19} and suggested earlier in \cite{Bev11}.  There are
furthermore possible $\gamma\gamma$ invariant mass structures at 10-15
MeV and 38 MeV in $pp$, and $\pi^- p$ reactions in COMPASS experiments
\cite{Ber11,Ber14,Bev12,Sch11,Sch12,Ber12,Bev20}.  Future
investigations in the region of low $\gamma \gamma$ invariant masses
will provide additional tests to confirm or refute the concept of the
open string QED mesons.

It is instructive to re-examine the theoretical basis for the possible
occurrence of such open string $q\bar q$ ~``QED mesons'' as proposed
in \cite{Won10}.  Because a quark $q$ and an antiquark $\bar q$ cannot
be isolated, the intrinsic motion of a composite $q\bar q$ system in
the lowest-energy states in 3+1 dimensions lies predominantly in 1+1
dimensions, as in an open string with the quark and the antiquark at
its two ends.  The approximate validity for the open string
description for the lowest-energy $q\bar q$ systems is theoretically
supported by the dual-resonance model \cite{Ven68}, Nambu and Goto
meson string model \cite{Nam70,Got71}, 'tHooft's two-dimensional meson
model \cite{tho74a,tho74b}, the classical yo-yo string model and the
Lund model \cite{Art74,And83}, the 2D inside-outside cascade model of
Bjorken \cite{Bjo73}, Casher, Kogut, and Susskind \cite{Cas74}, and
lattice gauge theories \cite{Hua88,Bal05,Cos17}.  The open
string description of a flux tube in hadron production at high
energies is experimentally supported by the limiting average
transverse momentum and a rapidity plateau
\cite{Cas74,Bjo73,Won91,Won94,Won09,Gat92} in high-energy $e^+$-$e^-$
annihilations \cite{Aih88,Hof88,Pet88,Abe99,Abr99} and $pp$ collisions
\cite{Yan08}.  To study approximately the lowest-energy bound states
of $q \bar q$ systems with light quarks, it is reasonable to truncate
the gauge field theories from 3+1 dimensions to 1+1 dimensions by
idealizing the three-dimensional flux tube as a structureless
one-dimensional open string, with the information on the structure of
the flux tube stored into the coupling constant of the interaction in
the lower 1+1 dimensions.  Whatever deviations from such a truncation
can be considered as perturbations.  The approximate validity of such
a truncation will need to be tested by confronting its theoretical
results with experiment.

In 1+1 dimensions, Schwinger already showed that a massless fermion
and an antifermion interacting with a gauge interaction give rise to a
bound boson \cite{Sch62,Sch63}.  If one identifies Schwinger's
massless fermion and antifermion as a light quark and a light
antiquark, one will reach the conclusion that a gauge interaction
between the quark and the antiquark in 1+1 dimensions 
leads to a confined and bound boson state with
a mass $m$, related to the gauge field coupling constant $g_{_{\rm
    2D}}$ by \cite{Sch62,Sch63}
\begin{eqnarray}
m^2=\frac{g_{_{\rm 2D}}^2} {{\pi}},
\label{eq1}
\end{eqnarray}    
which  shows that the mass $m$ of the bound boson increases as the 
strength $g_{\rm 2D}$ of the interaction increases, whether it be the
QED or the QCD interaction{\footnote{For a pedagogical derivation of
Schwinger's result of Eq.\ (\ref{eq1}), see for example, Chapter 6 of
\cite{Won94}.  For recent generalizations and extensions of the
Schwinger model, see \cite{Geo19,Geo19a,Geo20}.}. 

We need an important relationship to ensure that the boson mass
calculated in the lower 1+1 dimensions can appropriately represent the
mass of a physical boson in 3+1 dimensions.  In the physical world of
3+1 dimensions, the one-dimensional $q\bar q$ open string without a
structure is in fact an idealization of a flux tube with a transverse
radius $R_T$.  The boson masses calculated in 1+1 dimensions can
represent physical boson masses, when the structure of the flux tube
is properly taken into account.  Upon considering the structure of the
flux tube in the physical 3+1 dimensions, we find that the coupling
constant $g_{\rm 2D}$ in lower 1+1 dimensions is related to the
physical coupling constants $g_{\4d}$ in 3+1 dimensions by
\cite{Won09,Won10,Kos12}
\begin{eqnarray}
(g_{\2d})^2=\frac{1}{\pi
    R_T^2}(g_{\4d})^2=\frac{4\alpha_{\4d}}{R_T^2},
\label{12}
\end{eqnarray}
whose qualitative consistency can be checked by dimensional analysis.
Thus, when the dynamics in the higher dimensional 3+1 space-time is
approximated as dynamics in the lower 1+1 dimensions, information on
the flux tube structure is stored in the multiplicative conversion
factor $1/\pi R_T^2$ in the above equation that relates the physical
coupling constant square $(g_{4D})^2$ in 3+1 dimensions to the new
coupling constant square $ (g_{2D})^2$ in 1+1 dimensions.  As a
consequence, there is no loss of the relevant physical information.
The boson mass $m$ determined in 1+1 dimensions is the physical mass
related to the physical coupling constant
$\alpha_{\4d}$=$(g_{\4d})^2/4\pi$ and the flux tube radius $R_T$ by
\begin{eqnarray}
m^2= \frac{4\alpha_{\4d} }{\pi R_T^2}.
\end{eqnarray}
With 
$\alpha_{\4d}^{\qed}$\!=$\alpha_{{}_{\rm QED}}$=1/137,
$\alpha_{\4d}^{\qcd}$\!=$\alpha_s $$\sim$0.6 from hadron
spectroscopy \cite{Won00,Won01,Bal08,Deu16},
 and
$R_T$$\sim$0.4 fm from lattice QCD calculations \cite{Cos17} and 
$\langle p_T^2 \rangle $ of produced hadrons in high-energy $e^+e^-$
annihilations \cite{Pet88}, we estimate the masses of the open string
QCD and QCD mesons to be
\begin{eqnarray}
m^{\qcd}\sim 431{\rm ~ MeV}, ~~~ {\rm and}~~ m^{\qed}\sim 47 {\rm
  ~MeV}.
\label{14}
\end{eqnarray}
The above mass scales provide an encouraging guide for the present
task of a quantitative description of the QCD and QED mesons as open
strings, using QCD and QED gauge field theories in 1+ 1 dimensions.
Of course, the approximate validity of such a truncation of the $q\bar
q$ systems from 3+1 dimensions to 1+1 dimensions must be tested by
direct confrontation with experimental data, as will be carried out in
the next section.

\section{Open string  QCD and QED states  of $q\bar q$ systems} 

\subsection{Quarks and antiquarks  interacting with the QCD and
  QED interactions}\label{suba}

A quark and an antiquark carry color and electric charges.  They interact
with the QCD and the QED interactions, which are independent
interactions with different gauge symmetries, commutation properties,
and coupling constants.  The QED interaction is a U(1) gauge
interaction, whereas the QCD interaction is an SU(3) gauge
interaction.  They possess different generators and give rise to bound
boson states of quarks and antiquarks at different state energies, as
the order-of-magnitude estimates in (\ref{14}) indicate.

We would like to review and extend our earlier work \cite{Won10} on
the $q\bar q$ bound states with QCD and QED interactions in a single
framework.  We wish to extend our considerations from two flavors to
three flavors for the QCD interaction so that $\pi^0$, $\eta$, and
$\eta'$ can be adequately described.  A successful description of
these hadrons as open string $q\bar q$ QCD mesons will lend support
for its theoretical extrapolation into the unknown sector of $q\bar q$
QED mesons.

Accordingly, we introduce an enlarged U$(3)$ group that is the union
of the color SU(3) QCD subgroup and the electromagnetic U(1) QED
subgroup \cite{Won10}.  The generator $t^0$ for the U$(1)$ subgroup is
\begin{eqnarray}
t^0=\frac{1}{\sqrt{6}} \left ( \begin{matrix} 1 & 0 & 0 \cr 0 & 1 & 0
  \cr 0 & 0 & 1
        \end{matrix}   \right ),
\end{eqnarray}
which adds on to the eight standard generators of the SU$(3)$
subgroup, $\{t^1,...t^8\}$, to form the nine generators of the U(3)
group.  They satisfy $2 \,{\rm tr}\{ t^a t^ b \} = \delta^{ab} {~~\rm
  for~~} a,b=0,1,..,8$. The two subgroups of U(3) differ in their
coupling constants and communicative properties.  We consider quarks
with $N_f^\lambda$ number of flavors where $f$=$u,d,s$=1,2,3 is the
flavor label, and the superscript  
$\lambda$  is the 
interaction label with $\lambda$=0
for QED and $\lambda$=1 for QCD.  Because of the mass scale of
(\ref{14}), we have $m^{\qcd} \gg \{m_u, m_d, m_s\}$ and
$m^{\qed} $$\gg $$\{m_u, m_d\}$, and we can choose
$N_f^{\qed}$\!=$N_f^0$=2 and $N_f^{\qcd}$\!=$N_f^1$=3.

We start with 3+1 dimensional space-time $x^\mu$, with $\mu$=0,1,2,3.
The dynamical variables are the U(3) gauge fields $A_\mu$=$\sum_a
A_\mu^a t^a $ and the quark fields $\psi_f^i$ where $i$ is the color
index with $i$=1,2,3.  We use the summation convention over repeated
indices except when the summation symbols are needed to avoid
ambiguities.  For brevity of notations, the indices $\{a$,$f\}$ and
the superscript interaction labels $\{\lambda$, QCD, QED\} in various
quantities are often implicitly understood except when they are needed
to resolve ambiguities.  The coupling constants $g_f^a$ are given
explicitly by
\begin{subequations}
\begin{eqnarray}
\label{qedcc} 
&&\hspace{-0.9cm}g_u^0\!=\!-Q_u\,g_{\4d}^{\qed}\!\!\!\!,\!
~~~g_d^0\!=\!-Q_d\,g_{\4d}^{\qed} {\rm~~~~for~QED},
\\ &&\hspace{-0.9cm}g_{\{u,d,s\}}^{\{1,..,8\}}=Q_{\{u,d,s\}}^{{}^{\qcd}}\,g_{\4d}^\qcd
   {\rm~~~~~~~~~~~~~~for~QCD},
\label{qcdcc}
\end{eqnarray}
\end{subequations}
where we have introduced the charge numbers $Q_u^{\qed} $\!\!\!=2/3,
$Q_d^{\qed}$\!\!\!=$-$1/3, $Q_u^{\qcd}$\!\!\!=$~Q_d^{\qcd}
$\!\!\!=$~Q_s^{\qcd} $\!\!\!=~1.  The Lagrangian density for the
system is
\begin{subequations}
\begin{eqnarray}
{\cal L}&&=\bar \psi (i \sD)\psi - \frac{1}{4}F_{\mu \nu} F^{\mu \nu}-
m \bar \psi \psi, \\
 \hspace*{-2.0cm}\text{where}~~~~~~~~~~~~~~~~~~~~~~~~~ ~~~i\sD &&=
 \gamma^\mu ( i\sd~ + g A_\mu), \\
\label{F2}
F_{\mu \nu} &&= \partial_\mu A_\nu - \partial_\nu A_\mu -i g [A_\mu,
  A_\nu], ~~~~~~F_{\mu \nu}=F_{\mu \nu}^a t^a.
\end{eqnarray}
\end{subequations}
The equation of motion for the gauge field $A_\mu$ is
\begin{eqnarray}
\label{Max4}
D_\mu F^{\mu \nu}&& = \partial_\mu F^{\mu \nu} -i g [A_\mu,
  F^{\mu\nu}] = g j^{\nu}, ~~~~~ j^\nu= j^{\nu \, a} t^a, ~~~~j^{\nu
  \,a} =2 ~{\rm tr}~ {\bar \psi}_f \gamma^\nu t^a \psi_f.
\end{eqnarray} 

As a quark and an antiquark cannot be isolated, the intrinsic motion
of the quark and the antiquark in the lowest-energy $q\bar q$ systems
in 3+1 dimensions lies predominantly in 1+1 dimensions, as discussed
in string models of mesons
\cite{Ven68,Nam70,Got71,tho74a,tho74b,Art74,And83,Bjo73,Cas74,Col75,Col76,
  Won09,Won10,Won11,Kos12}.  We shall therefore approximate the gauge
field theory in 3+1 dimensions by the gauge field theory in 1+1
dimensions where the coupling constant $g$ will be implicitly taken to
be $g_{\2d}$.  It is necessary to keep in mind that the information on
the structure of flux tube radius $R_T$ is stored in the
multiplicative factor $1/(\sqrt{\pi} R_T)$ that converts the physical
coupling constant $g_{\4d}$ to the new coupling $g_{\2d}$ in the lower
dimensions as given by (\ref{12}).

\subsection{Bosonization of QCD and QED for $q\bar q$ systems in 1+1 dimensions}

We wish to search for bound states arising from the interaction of the
color and electric charges of the quarks and antiquarks in QCD and QED
in the strong coupling limit in 1+1 dimensions.  The bound states can
be searched by the method of bosonization in which the stability of
the boson states can be examined by the values of the square of the
boson mass, with residual sine-Gordon interactions that depend on the
quark mass \cite{Col76}-\cite{Kov11}.

The U(3) gauge interactions under consideration contain the
non-Abelian color SU(3) interactions.  Consequently the bosonization
of the color degrees of freedom should be carried out according to the
method of non-Abelian bosonization which preserves the gauge group
symmetry \cite{Wit84}.

While we use non-Abelian bosonization for the U(3) gauge interactions,
we shall follow Coleman to treat the flavor degrees of freedom as
independent degrees of freedom \cite{Col76,Ell92,Fri93,Nag09,Kov11}.
This involves keeping the flavor labels in the bosonization without
using the flavor group symmetry.  Although such bosonization in the
flavor sector obscures the isospin and other flavor symmetry in QCD,
the QCD isospin and other flavor symmetry are still present.  They can
be recovered by complicated non-linear general transformations
\cite{Col76,Hal75}, or by using explicit  physical $q\bar q$ multi-flavor eigenstates 
as a linear combination of flavor components.  
As such a bosonization method is simple only for
neutral $q\bar q$ systems with isospin component $I_3$=0, we shall
limit our attention to such systems.  We shall study only $q\bar q$
systems with total spin $S$=0.

As in any method of bosonization, the non-Abelian method will succeed
for systems that contain stable and bound boson states with relatively
weak residual interactions.  Thus, not all the degrees of freedom
available to the bosonization technique will lead to good boson states
with these desirable properties. For example, some of the bosonization
degrees of freedom in color SU(3) may correspond to bosonic
excitations into colored objects of two-fermion complexes and may not
give rise to stable bosons.  It is important to judiciously search for
those boson degrees of freedom that will eventually lead to stable and
bound bosons.  Keeping this perspective in our mind, we can examine
the non-Abelian bosonization of the system under the U(3)
interactions.  The non-Abelian bosonization program consists of
introducing boson fields $\phi^a$ to describe an element $u$ of the
U(3) group and showing subsequently that these boson fields lead to
stable bosons with finite or zero masses.

In the non-Abelian bosonization, the current $j_\pm$ in the light-cone
coordinates, $x^{\pm}$=$(x^0 \pm x^3)/\sqrt{2}$, is bosonized as
\cite{Wit84}
\begin{subequations}
\label{jj}
\begin{eqnarray}
\label{jja}
j_+ & = & ~~(i/2\pi) u^{-1} (\partial_+ u),\\
\label{jjb}
j_- & = & -(i/2\pi) (\partial_- u) u^{-1}.
\end{eqnarray}
\end{subequations}
An element of the U(1) subgroup of the U(3) group can be represented
by the boson field $\phi^0$
\begin{eqnarray}
u = \exp\{ i 2 \sqrt{\pi} \phi^0 t^0 \}.
\end{eqnarray}  
Such a bosonization poses no problem as it is an Abelian subgroup.  It
will lead to a stable boson as in Schwinger's QED2.

To carry out the bosonization of the color SU(3) subgroup, we need to
introduce boson fields to describe an element $u$ of SU(3).  There are
eight $t^a$ generators which provide eight degrees of freedom.  We
may naively think that for the non-Abelian bosonization of SU(3), we
should introduce eight boson fields $\phi^a$ to describe $u$ by
\begin{eqnarray}
u = \exp\{ i 2 \sqrt{\pi} \sum_{a=1}^8 \phi^a t^a\}.
\end{eqnarray}
However, a general variation of the element $\delta u /\delta x^\pm$
will lead to quantities that in general do not commute with $u$ and
$u^{-1}$, resulting in $j_\pm$ currents in Eqs.\ ({\ref{jj}) that are
complicated non-linear admixtures of the boson fields $\phi^a$.  It
will be difficult to look for stable boson states with these
currents.

We can guide ourselves to a situation that has a greater chance of
finding stable bosons by examining the bosonization problem from a
different viewpoint.  We can pick a unit generator
${\tau}^1$=$\sum_{a=1}^8 n_a t^a$ with $n_a$=$2{\rm tr}(\tau^1 t^a)$
oriented in any direction of the eight-dimensional generator space and
we can describe an SU(3) group element $u$ by an amplitude $\phi^1$
and the unit vector $\tau^1$,
\begin{eqnarray}
\label{choi}
u = \exp\{ i 2 \sqrt{\pi} \phi^1 \tau^1 \}.
\end{eqnarray}
The boson field $\phi^1$ describes one degree of freedom, and the
direction cosines $\{n^a,a=1,..,8\}$ of the unit vector $\tau^1$
describe the other seven degrees of freedom.  A variation of the
amplitude $\phi^1$ in $u$ while keeping the unit vector orientation
fixed will lead to a variation of $\delta u/\delta x^\pm $ that will
commute with $u$ and $u^{-1}$ in the bosonization formula (\ref{jj}),
as in the case with an Abelian group element.  It will lead to simple
currents and stable QCD bosons with well defined masses, which will
need to be consistent with experimental QCD meson data.  On the other
hand, a variation of $\delta u/\delta x^\pm $ in any of the other
seven orientation angles of the unit vector $\tau^1$ will lead to
$\delta u/\delta x^\pm $ quantities along other $t^a$ directions with
$a$=$\{1,...,8\}$.  These variations of $\delta u/\delta x^\pm $ will
not in general commute with $u$ or $u^{-1}$.  They will lead to
$j_\pm$ currents that are complicated non-linear functions of the
eight degrees of freedom. We are therefore well advised to search for
stable bosons by varying only the amplitude of the $\phi^1$ field,
keeping the orientation of the unit vector fixed, and forgoing the
other seven orientation degrees of freedom.  For the U(3) group, there
is in addition the group element $u = \exp\{ i 2 \sqrt{\pi} \phi^0 t^0
\}$ from the QED U(1) subgroup.  Combining both U(1) and SU(3)
subgroups, we can represent an element $u$ of the U(3) group by
$\phi^0$ from QED and $\phi^1$ from QCD as \cite{Won10}
\begin{eqnarray}
\label{uua}
u = \exp\{ i 2 \sqrt{\pi} \phi^0 \tau^0 + i 2 \sqrt{\pi} \phi^1
\tau^1\},
\label{uuu}
\end{eqnarray}
where we have re-labeled $t^0$ as $\tau^0$.  The superscripts
$\lambda$=$\{ 0,1\}$ on the right hand side of the above equation is
the interaction label with $\lambda$=$\{ 0,1\}$ for QED and QCD,
respectively, and $2{\rm tr}(\tau^\lambda\tau^{\lambda'}
)$=$\delta^{\lambda{\lambda'}}$.  When we write out the flavor index
explicitly, we have
\begin{eqnarray}
\label{uuaf}
u_f = \exp\{ i 2 \sqrt{\pi} \phi_f^0 \tau^0 + i 2 \sqrt{\pi} \phi_f^1
\tau^1\}.
\end{eqnarray}
From (\ref{jja}) and (\ref{jjb}), we obtain
\begin{subequations}
\begin{eqnarray}
j_{f\pm} &=& \mp \frac{1}{\sqrt{\pi}} \left [ (\partial _\pm\phi_f^0)
  \tau^0 + (\partial _\pm\phi_f^1)\tau^1\right ]~~~~~~~~~{\rm
  when~all} ~Q_f^\lambda=1, \\ &=& \mp \frac{1}{\sqrt{\pi}} \left [
  Q_f^0(\partial _\pm\phi_f^0) \tau^0 + Q_f^1 (\partial
  _\pm\phi_f^1)\tau^1\right ]~\text{when we include charge number}
~Q_f^\lambda.~~~~~~~~~~
\end{eqnarray}
\end{subequations}
The Maxwell equation in light-cone coordinates is
\begin{eqnarray}
\partial _\mu \partial ^\mu A^\pm - \partial _ \pm \partial _\mu A^\mu
&=& g j^ \pm .
\end{eqnarray}
We shall use the Lorenz gauge
\begin{eqnarray}
\partial _\mu A^\mu =0,
\end{eqnarray}
then the solution of the gauge field is
\begin{eqnarray}
A^\pm &=& \frac{g}{ 2\partial _+ \partial _-} j^ \pm.
\end{eqnarray}
Interaction energy ${ H}_{\rm int}$ is
\begin{eqnarray}
{ H}_{\rm int} &&= \frac{g}{2}\int dx^+dx^-~2\,{\rm tr}\, (j \cdot A
)=\frac{g}{2}\int dx^+dx^- ~2\,{\rm tr}\, ( j^+ A^- + j^- A^+)
\nonumber\\ &&=\frac{g}{2}\int dx^+dx^- ~2\,{\rm tr}\,\left ( j^+
\frac{g}{ 2\partial _+ \partial _-} j^- + j^- \frac{g}{ 2\partial _+
  \partial _-} j^+ \right ).
\end{eqnarray}
We integrate by parts, include the charge numbers and the interaction
dependency of the coupling constant, $g_{\rm 2D}^\lambda$=$g^\lambda$,
and we obtain the contribution to the Hamiltonian density from the
confining interaction between the constituents, $H_{\rm int}=\int dx^+
dx^- {\cal H}_{\rm int}(\phi_f^\lambda)$,
\begin{eqnarray}
{\cal H}_{\rm int}(\phi_f^\lambda) &=&\frac{1}{2} \biggl [
  \frac{(g_{\2d}^0)^2}{\pi} (\sum_f^{N_f} Q_{f}^0\phi_f ^0 )^2 +
  \frac{(g_{\2d}^1)^2}{\pi} (\sum_f^{N_f} Q_{f}^1\phi_f^1)^2 \biggr ],
\label{216}
\end{eqnarray}
which matches the results of \cite{Col76,Nag09}.  For the mass
bi-linear term, we follow Coleman \cite{Col76} and Witten \cite{Wit84}
and bosonized it as
\begin{eqnarray}
m_f :\bar \psi_f \psi_f :&& \to (-\frac{e^\gamma}{2\pi} ) \mu m_f
~2{\rm tr} \left ( \frac{u_f + u_f^{-1}}{2} \right ), \nonumber\\ &&=
(-\frac{e^\gamma}{2\pi} ) \mu m_f ~2{\rm tr} \cos \left (2 \sqrt{\pi}
\phi_f ^0\tau^0 + 2\sqrt{\pi} \phi_f^1\tau^1 \right ),
\end{eqnarray}
where $\gamma=0.5772$ is the Euler constant, and $\mu$ is an unknown
mass scale that arises from the bosonization of the scalar density
${\bar \psi} \psi$ and is solution-dependent \cite{Col76}.  When we
sum over flavors, we get the contribution to the Hamiltonian density
from quark masses, 
\begin{eqnarray}
H_{\rm m}=\int dx^+ dx^- {\cal H}_{\rm  m}(\phi_f^\lambda),
\end{eqnarray}
where
\begin{subequations}
\begin{eqnarray}
{\cal H}_{\rm m} ( \phi_f^\lambda)\!&=&\!e^\gamma \mu \sum_f\! m_f \left [
  (\phi_f^0 )^2 + (\phi_f^1)^2 \! + \!...\right ]~~~\text{when $\mu$  is independent of interaction},~~~~~~~~~
\\
    &=&\!e^\gamma \sum_f\! m_f \left [ \mu^0 (\phi_f^0 )^2\! + \!\mu^1 (\phi_f^1)^2+ ...\right ]~\text{when \!$\mu$ depends on  interaction}.~~~~~~~~
\label{218} 
\end{eqnarray}
\end{subequations}
Finally, for the kinematic term, we bosonize it as \cite{Wit84,Gep85}}
\begin{eqnarray}
: \bar \psi i \sd \psi : ~~~~ \to~~~ \frac{1}{8\pi} \biggl  \{ 2~{\rm tr} \left [
    \partial_ \mu u )~(\partial ^\mu u ^{-1})\right ] \biggr  \} + n\Gamma,
\end{eqnarray}
where $n\Gamma$ is the Wess-Zumino term which vanishes for $u$ of
(\ref{uuu}) containing commuting elements $\tau^0$ and $\tau^1$.  We
get the kinematic contribution
 \begin{eqnarray}
&H_{_{\rm kin}}=\int dx^0 dx^1 {\cal H}_{_{\rm kin}} =\int dx^0 dx^1
   \sum_f \frac{1}{8 \pi} \biggl  \{~2\,{\rm tr} \left [ \partial_ \mu u_f
     )~(\partial ^\mu u_f ^{-1})\right ] \biggr  \}, 
\end{eqnarray}
where
\begin{eqnarray}
&{\cal
     H}_{_{\rm kin}}(\phi_f^\lambda)= \frac{1 }{2}\sum_f \left [
     \partial _\mu\phi_f ^0\partial ^\mu \phi_f ^0 + \partial
     _\mu\phi_f ^1\partial ^\mu \phi_f ^1\right ]= \frac{1 }{2}
   \sum_\lambda \sum_f \left [ (\Pi_f^\lambda)^2 + (\partial _x\phi_f
     ^\lambda)^2 \right ],
\label{220}
\end{eqnarray}
and $\Pi_f^\lambda$ is the momentum conjugate to
$\phi_f^\lambda$. The total Hamiltonian density in terms of
$\phi_f^\lambda$ is
\vspace*{-0.3cm}
\begin{eqnarray}
{\cal H}(\phi_f^\lambda)= {\cal H}_{_{\rm kin}}(\phi_f^\lambda)+{\cal
  H}_{\rm int}(\phi_f^\lambda)+{\cal H}_{\rm m}(\phi_f^\lambda),
\label{221}
\end{eqnarray}
where $ {\cal H}_{_{\rm kin}}(\phi_f^\lambda)$, ${\cal H}_{\rm
  int}(\phi_f^\lambda)$, ${\cal H}_{\rm m}(\phi_f^\lambda)$ are given
by Eqs.\ (\ref{220}), (\ref{216}), and (\ref{218}) respectively.

\subsection{Orthogonal transformation to  $q\bar q$\, flavor eigenstates}

We consider $q\bar q$ systems with dynamical flavor symmetry that lead
to flavor eigenstates as a linear combination of states with different
flavor amplitudes.  Such eigenstates arise from additional
considerations of isospin invariance, SU(3) flavor symmetry, and
configuration mixing.  As a result of such considerations in flavor
symmetry and configuration mixing, the physical eigenstates $\Phi_i$
can be quite generally related to various flavor components $\phi_f$
by a linear orthogonal transformation as
\begin{eqnarray}
\Phi_i^\lambda=\sum_ f D_{if}^\lambda \phi_f ^\lambda.
\end{eqnarray}
The orthogonal transformation matrix $D_{if}^\lambda$ obeys 
$(D^\lambda)^{-1}=(D^\lambda)^\dagger$ with
$((D^\lambda)^\dagger)_{fi}$=$D_{if}^\lambda$.  The inverse transformation is
\begin{eqnarray}
\phi_f^\lambda=\sum_ i D_{if}^\lambda \Phi_i^\lambda.
\end{eqnarray}
Upon substituting the above equation into (\ref{221}), we obtain the
total Hamiltonian density in terms of the physical flavor state
$\Phi_i^\lambda$ as
\begin{eqnarray}
{\cal H}(\Phi_f^\lambda)= [{\cal H}_{_{\rm kin}}(\Phi_i^\lambda)+{\cal
    H}_{\rm int}(\Phi_i^\lambda)+{\cal H}_{\rm m}(\Phi_i^\lambda)],
\end{eqnarray}
\begin{subequations}
\begin{eqnarray}~~
\hspace*{-2.5cm}\text{where}~~~~~~~~{\cal H}_{_{\rm kin}}
(\Phi_i^\lambda) &&= \frac{1 }{2} \sum_\lambda \sum_i \left [\partial
  _\mu\Phi_i ^\lambda\partial ^\mu \Phi_i ^\lambda \right ] = \frac{1
}{2} \sum_\lambda \sum_i \left [ (\Pi_i^\lambda)^2+ (\partial
  _x\Phi_i^\lambda)^2 \right ], \\ {\cal H}_{\rm int}(\Phi_i^\lambda)
&&=\frac{1}{2}\biggl [ \sum_\lambda \frac{(g_{\2d}^\lambda)^2}{\pi}
  (\sum_f Q_{f}^\lambda\sum_i D_{if}^\lambda \Phi_i ^\lambda )^2 \biggr ],
\label{225b}
\\ {\cal H}_{\rm m}(\Phi_i^\lambda)&&=e^\gamma \sum_f m_f \left [
  \sum_\lambda \mu^\lambda (\sum_i D_{if}^\lambda
  \Phi_i^\lambda )^2 \right ],
\label{225c}
\end{eqnarray}
\end{subequations}
where $\Pi_i^\lambda$ is the momentum conjugate to $\Phi_i^\lambda$.
We can get the boson mass $m_i^\lambda$ of the physical state
$\Phi_i^\lambda$ by expanding the potential energy term, ${\cal
  H}_{\rm int}(\Phi_i^\lambda)+{\cal H}_{\rm m}(\Phi_i^\lambda)$,
about the potential minimum located at $\Phi_i^\lambda=0$, up to the
second power in $(\Phi_i^\lambda)^2$, as
\begin{eqnarray}
{\cal H}(\Phi_i^\lambda) &&= \sum_\lambda \sum_i\left [ \frac{1 }{2}
  (\Pi_i ^\lambda)^2 +\frac{1 }{2} (\partial _x \Phi_i ^\lambda)^2
  +\frac{1}{2} (m_{i}^\lambda)^2 (\Phi_i ^\lambda)^2 \right ]+ ...,
\end{eqnarray}
\begin{eqnarray}
\hspace*{-3.0cm}\text{where}~~~~~~~~~~~~~~~~&&(m_{i}^\lambda)^2
=\biggl [ \frac{\partial^2}{\partial (\Phi_{i}^\lambda)^2} [{\cal
      H}_{\rm int} (\{\Phi_i^\lambda\})+ {\cal H}_{\rm
      m}(\{\Phi_i^\lambda\})]\biggr
]_{\Phi_0^\lambda,\Phi_1^\lambda=0} .
\end{eqnarray}
From Eqs.\ (\ref{225b}) and (\ref{225c}), we find the squared mass
$(m_i^\lambda)^2$ for the state $\Phi_i^\lambda$ of interaction $\lambda$
to be
\begin{eqnarray}
(m_{i}^\lambda)^2&&=\frac{(g_{\2d}^\lambda)^2}{\pi} \left [
    \sum_f^{N_f} Q_{f}^\lambda D_{if}^\lambda \right ] ^2 + e^\gamma
  \sum_f^{N_f} m_f \mu^\lambda(D_{if}^\lambda)^2.
\label{230}
\end{eqnarray}
This mass formula includes the mixing of the configurations, and is
applicable to QCD with three flavors.  It is an improved and more
general extension of the earlier mass formula in \cite{Won10}.  It
should be reminded that the coupling constant $g_{\2d} ^\lambda$ in
1+1 dimensions above is related to the coupling constant
$g_{\4d}^\lambda$ in 3+1 dimensions by the flux tube radius in
(\ref{12}).  A positive definite value of $(m_{i}^\lambda)^2$, which
is ensured by the positive quantities on the right hand side of
(\ref{230}), indicates that the boson from such an interacting system
of $q$ and $\bar q$ are stable bosons.

The two terms on the right hand side of (\ref{230}) receive
contributions from different physical sources.  The first term, the
``massless quark limit'' or alternatively the ``confining
interaction term'', arises from the confining interaction between the
quark and the antiquark.  The second term arises from quark masses and
the quark condensate, $\langle \sum_f \bar \psi_f \psi_f\rangle $.  It
can be called the ``quark mass term'' or the ``quark condensate
term''.  If one labels the square root of the first term in
(\ref{230}) as the confining interaction mass and the square root of
the second term as the condensate mass, then the hadron mass obeys a
Pythagoras theorem with the hadron mass as the hypotenuse and the
confining interaction mass and the condensate mass as two sides of a
right triangle.

\subsection{Open string description of the QCD mesons \label{sec24} }

QCD has an approximate SU(3)$_L\times$ SU(3)$_R$ chiral symmetry and
also an approximate flavor U(3)$\times$U(3) symmetry.  If the axial
symmetry is realized as the Goldstone mode as a result of the
spontaneous chiral symmetry breaking, then one would naively expect
the singlet isoscalar $\eta'$ particle to have a mass comparable to
the pion mass.  Experimentally, there is the U$_A$(1) anomaly
\cite{Kog74a,Kog75a,Kog75b,Wei79,Wit79} in which the $\eta'$ mass of
957.8 MeV is so much higher than the $\pi$ mass.  On the basis of the
Schwinger model, Kogut, Susskind, and Sinclair
\cite{Kog74a,Kog75a,Kog75b} suggested that such a U$_A$(1) anomaly
arises from the long-range confinement between the quark and the
antiquark, as the $\eta'$ acquires a large mass from the long-range
confining interaction between a quark and an antiquark.  The long
range gauge interaction affects not only $\eta'$ mass but also the
other pseudoscalar $\pi^0$, and $\eta$ masses, and there are
furthermore the effects of quark rest masses, and the configuration
mixing between $\eta$ and $\eta'$.  We would like to show that when
these effects are properly taken into account, the pseudoscalar
particles $\pi^0$, $\eta$, and $\eta'$ can indeed be adequately
described as open string QCD mesons.

Accordingly, in this subsection (\ref{sec24}) to study QCD mesons, we
restrict ourselves to the QCD interaction with the interaction label
superscript set implicitly to $\lambda$=1 for QCD.  Equation (\ref{14})
indicates that the mass scale $m^{\qcd}\!\!\!\!\sim$ 431 MeV $ \gg
m_u, m_d, m_s$.  It is necessary to include $u$, $d$, and $s$ quarks
with $N_f $=3 in the analysis of open strings QCD mesons.

We denote $\phi_1=|u \bar u \rangle$, $\phi_2=|d \bar d \rangle$, and
$\phi_3=|s \bar s \rangle$, and assume the standard quark model
description of $|\pi^0\rangle$, $|\eta\rangle$, and $|\eta'\rangle$ in
terms of flavor octet and flavor singlet states, with the mixing of
the $|\eta\rangle$ and $|\eta'\rangle$ represented by a mixing angle
$\theta_P$ \cite{PDG19}.  The physical states of $|\pi^0\rangle$,
$|\eta\rangle$, and $|\eta'\rangle$ can be represented in terms of the
flavor states $\phi_1$, $\phi_2$ and $\phi_3$ by
\begin{subequations}
\begin{eqnarray}
|\pi^0\rangle&=&\Phi_1=\frac{\phi_1-\phi_2}{\sqrt{2}},
\\ |\eta~\rangle&=&\Phi_2= |\eta_8\rangle \cos \theta_P- |
\eta_0\rangle \sin \theta _P,
\label{229b}
\\ |\eta'\rangle&=&\Phi_3= |\eta_8 \rangle \sin \theta_P+|
\eta_0\rangle \cos \theta _P,
\label{229c}
\end{eqnarray}
where
\begin{eqnarray}
&&|\eta_8\rangle
=\frac{\phi_1+\phi_2-2\phi_3}{\sqrt{6}}, \\ & &\,|\eta_0\rangle
=\frac{\sqrt{2}(\phi_1+\phi_2+\phi_3) }{\sqrt{6}}.
\end{eqnarray}
\end{subequations}
The physical states $\Phi_i$=$\sum_f D_{if}\phi_f$ and the flavor
component states $\phi_f$, are then related by
\begin{eqnarray}
\begin{pmatrix}
 \Phi_1\\ \Phi_2\\ \Phi_3
\end{pmatrix}
\!\!=\!\!\begin{pmatrix} \frac{1}{\sqrt{2} } & - \frac{1}{\sqrt{2} } &
0 \\ \frac{1} {\sqrt{6}} \{ \cos \theta_P \!-\!\sqrt{2} ~ \sin
\theta_P\} & \frac{1} {\sqrt{6}} \{ \cos \theta_P \!-\!\sqrt{2} ~ \sin
\theta_P\} & \frac{1} {\sqrt{6}}\{-2\cos \theta_P \!-\! \sqrt{2}\sin
\theta_P\} \\ \frac{1}{\sqrt{6}}\{\sin \theta_P \!+\! \sqrt{2}\cos
\theta_P\} & \frac{1}{\sqrt{6}}\{\sin \theta_P \!+\! \sqrt{2}\cos
\theta_P\} & \frac{1}{\sqrt{6}}\{-2\sin \theta_P\! +\! \sqrt{2}\cos
\theta_P\} \\
\end{pmatrix}
\!\!
\begin{pmatrix}
 \phi_1\\ \phi_2\\ \phi_3
\end{pmatrix}\!\!,
\nonumber
\end{eqnarray}
with the inverse relation $\phi_f= \sum_{i=1}^3 D_{if}\Phi_i$,
\begin{eqnarray}
\begin{pmatrix}
 \phi_1\\ \phi_2\\ \phi_3
\end{pmatrix}=
\begin{pmatrix}
 \frac{1}{\sqrt{2} } & ~~~ \frac{1} {\sqrt{6}} \{ \cos \theta_P
 -\sqrt{2} ~ \sin \theta_P\}~~~ & \frac{1} {\sqrt{6}}\{ \sin \theta _P
 + \sqrt{2} ~\cos \theta _P \} \\ - \frac{1}{\sqrt{2} } & \frac{1}
 {\sqrt{6}}\{ \cos \theta_P - \sqrt{2} ~\sin \theta_P\} & \frac{1}
 {\sqrt{6}}\{ \sin \theta _P + \sqrt{2} ~\cos \theta _P \} \\ 0
 &\frac{1} {\sqrt{6}}\{-2\cos \theta_P \!-\! \sqrt{2}\sin \theta_P\} &
 \frac{1}{\sqrt{6}}\{-2\sin \theta_P\! +\! \sqrt{2}\cos \theta_P\} \\
\end{pmatrix}
\begin{pmatrix}
 \Phi_1\\ \Phi_2\\ \Phi_3
\end{pmatrix}.~~~~~~
\end{eqnarray}
With color charge $Q_{\{u,s,d\}}^{\qcd}$=1, the mass formula
(\ref{230})
gives
\vspace*{-0.2cm}
\begin{eqnarray}
m_i^2=(\sum_{f=1}^{N_f} D_{if})^2 \frac{4\alpha_s}{ \pi R_T^2} +
\sum_{f=1}^{N_f} m_f (D_{if})^2 e^\gamma \mu ^{\qcd} ,
\label{231}
\end{eqnarray}
yielding an effective color charge $Q_{i,{\rm eff}}$=$|\sum_{f=1}^3
(D_{if})|$.

For the pion state $\Phi_1$, we have $Q_{1,{\rm eff}}$=$|\sum_{f=1}^3
(D_{1f})|$=$|1/\sqrt{2}-1/\sqrt{2}|$=0. The first term of the
``massless quark  limit'' in the mass formula (\ref{231}) gives
a zero pion mass.  When the quark masses are taken into account, we
have $(D_{if})^2$=1/2.  The only contribution comes from the second
``quark condensate'' term in (\ref{231}).  The mass formula
(\ref{231}) for the pion then gives
\begin{eqnarray}
m_\pi^2= (m_u + m_d)\frac{ e^\gamma \mu^\qcd}{2},
\end{eqnarray}
which is in the same form as the Gell-Mann-Oakes-Renner relation
\cite{Gel68},
\begin{eqnarray}
m_\pi^2= (m_u + m_d) \frac{ |\langle 0|\bar q q |0 \rangle|}{F_\pi^2},
\label{233}
\end{eqnarray}
where $|\langle0| \bar q q |0\rangle|$ is the light $u$ and $d$
quark-antiquark condensate and $F_\pi$ is the pion decay constant
\cite{Wei95}.  Consequently, we can infer that the unknown mass scale
$\mu^\qcd$ in the bosonization formula for QCD has indeed the physical meaning of
the quark condensate.  We can therefore identify $\mu^\qcd$ in the
bosonization mass formula (\ref{231}) for QCD as
\begin{eqnarray}
\frac{e^\gamma \mu^\qcd}{2} = \frac{ |\langle 0|\bar q q|0
  \rangle|}{F_\pi^2}.
\label{236}
\end{eqnarray}
By such an identification and calibrating the pion mass to be the
experimental mass $m_\pi$, the mass formula (\ref{231}) for the
pseudoscalar QCD mesons can be re-written as
\begin{eqnarray}
m_i^2=(\sum_{f=1}^{N_f} D_{if})^2 \frac{4\alpha_s}{ \pi R_T^2} +
m_\pi^2 \sum_{f=1}^{N_f} \frac{m_f}{m_{ud}} (D_{if})^2,
\label{qcd}
\end{eqnarray}
where $m_{ud}=(m_u+m_d)/2$.  We are ready to test whether the $I_3$=0,
$S$=0 hadrons of $\pi^0$, $\eta$ and $\eta'$ can be appropriately
described as open strings in the 1+1 dimensional bosonization model.
For these QCD mesons, there is a wealth of information on the matrix
$D_{if}$ that describes the composition of the physical states in
terms of the flavor components, as represented by the mixing angle
$\theta_P$ between the flavor octet and flavor singlet components of
the SU(3) wave functions in $\eta$ and $\eta'$ in (\ref{229b}) and
(\ref{229c}). The ratio of the strange quark mass to the light $u$ and
$d$ quark masses that is needed in the above mass formula is also
known.  From the tabulation in PDG \cite{PDG19}, we find
$\theta_P=-24.5^o$ and ${m_s}/m_{ud}$= 27.3$_{-1.3}^{+0.7}$.  The only
free parameters left in the mass formula (\ref{qcd}) are the strong
interaction coupling constant $\alpha_s$ and the flux tube radius
$R_T$.

For the value of $\alpha_s$, previous works on the non-perturabtive
potential models use a value of $\alpha_s$ of the order of $0.4-0.6$
in hadron spectroscopy studies \cite{Won00,Won01,Bal08,Deu16}.
However, these potential models contain a linear confining
interaction, in addition to the one-gluon exchange interaction
involving $\alpha_s$.  In contrast, the present simplified 1+1
dimensional treatment uses only a single attractive gauge interaction,
involving $\alpha_s$ and playing dual roles.  We should be prepared to
allow for a larger value of the strong coupling constant $\alpha_s$ in
our case.  We find that $\alpha_s$=0.68 gives a reasonable description
of the masses of the mesons considered, and we can take the difference
between this $\alpha_s$ value and the $\alpha_s$ value of $0.6$ used
for the lowest meson masses in earlier hadron spectroscopy studies
\cite{Won00,Won01,Bal08,Deu16} as a measure of the degree of
uncertainties in $\alpha_s$, resulting in $\alpha_s$=0.68 $\pm$ 0.08.

For the value of $R_T$, lattice gauge calculations with dynamical
fermions give a flux tube root-mean-square-radius $R_T$=0.411 fm for a
quark-antiquark separation of 0.95 fm \cite{Cos17}.  The experimental
value of $ \langle p_T^2\rangle $ of produced hadrons ranges from 0.2
to 0.3 GeV$^2$ for $e^+$-$e^-$ annihilations at $\sqrt{s}$ from 29 GeV
to 90 GeV \cite{Pet88}, corresponding to a flux tube radius
$R_T$=$\hbar/\sqrt{\langle p_T^2\rangle }$ of 0.36 to 0.44 fm.  It is
reasonable to consider flux tube radius parameter to be $R_T=0.4 \pm
0.04$ fm.  This set of parameters of $\alpha_s$=0.68$\pm$0.08 and
$R_T$=0.40$\pm$0.04 fm give an adequate description of the $\pi^0$,
$\eta$ and $\eta'$ masses as shown in Table I.

\begin{table}[H]
\centering
\caption { Comparison of experimental and theoretical masses of
  neutral, $I_3$=0, and $S$=0 QCD and QED mesons obtained with the
  semi-empirical mass formula (\ref{qcd}) for QCD mesons and
  (\ref{qed}) for QED mesons, with $\alpha_{{}_{\rm QED}}$=1/137,
  $\alpha_s$=0.68$\pm$0.08, and $R_T$=0.40$\pm$0.04 fm.  }
\vspace*{0.2cm}
\begin{tabular}{|c|c|c|c|c|c|c|c|c|}
\cline{3-8} \multicolumn{2}{c|}{}& & & &Experimental& Semi-empirical
&Meson mass\\ \multicolumn{2}{c|}{}&I & S &$[I(J^\pi$)] & mass & mass
& in massless \\ \multicolumn{2}{c|}{}& & & & & formula & quark limit
\\ \multicolumn{2}{c|}{}& & & & (MeV) & (MeV) & (MeV) \\ 
\hline
QCD&$\pi^0$ & 1 & 0 &[1(0$^-$)] &\!\!134.9768$\pm$0.0005\!\!& 134.9$^\ddagger$ & 0 \\ 
\!\!meson\!\!& $\eta$ & 0 & 0 &[0(0$^-$)] &\!\!547.862$\pm$0.017\!\!&498.4$\pm$39.8~~ & 329.7$\pm$57.5~ \\ 
& $\eta'$ & 0 & 0 &[0(0$^-$)] & 957.78$\pm$0.06& 948.2$\pm$99.6~ &723.4$\pm$126.3 \\ \hline
QED&\!\!isoscalar\!\!& 0 & 0 &[0(0$^-$)] & & 17.9$\pm$1.5 & 11.2$\pm$1.3 \\ 
\!\!meson\!\!&\!\!isovector\!\!& 1 & 0 &[1(0$^-$)] & & 36.4$\pm$3.8&33.6$\pm$3.8 \\ 
\hline
Possible& X17 & & & (1$^+$)? &\!\!16.70$\pm$0.35$\pm$0.5$^\dagger$~~\!\!& & \\ 
QED& X17 & & & (0$^-$)?  &\!\!16.84$\pm$0.16$\pm$0.20$^\#$\!\!& & \\ 
meson& E38 & & &              ?  &  37.38$\pm$0.71$^\oplus$& & \\ 
\!\!candidates\!\!& E38 & & &              ?  &  40.89$\pm$0.91$^\ominus$& & \\ 
                               & E38 & & &              ?  &  39.71$\pm$0.71$^\otimes$& & \\ \hline
\end{tabular}
\vspace*{0.1cm}

\hspace*{-5.95cm}$^\ddagger$ Calibration mass~~~~~~~~~~~~~~~~~~~~~~~\\
$^\dagger$\,A. Krasznahorkay $et~al.$, Phys.Rev.Lett.116,042501(2016), $^8$Be$^*$ decay\\
\hspace*{-2.25cm}$^\#$A. Krasznahorkay $et~al.$, arxiv:1910.10459, $^4$He$^*$ decay~~~~\\
\hspace*{-0.70cm}$^\oplus$\,K. Abraamyan $et~al.$, EPJ Web Conf       204,08004(2019),$d$Cu$\to$$\gamma \gamma X$~\\
\hspace*{-0.70cm}$^\ominus$\,K. Abraamyan $et~al.$, EPJ Web Conf       204,08004(2019),$p$Cu$\to$$\gamma \gamma X$~\\
\hspace*{-0.70cm}$^\otimes$\,K. Abraamyan $et~al.$, EPJ Web Conf       204,08004(2019),~$d$C$\to$$\gamma \gamma X$~\\
\label{tb1}
\end{table}

From our comparison of the experimental and theoretical masses in
Table \ref{tb1}, we find that by using the method of bosonization and
including the confining interaction and the quark condensate, the mass
formula (\ref{qcd}) in the 1+1 dimensional open string model can
indeed describe adequately the masses of $\pi^0$, $\eta$, and $\eta'$,
approximately within the limits of the uncertainties of the
theory. The formulation can be used to extrapolate to the unknown
region of open string $q\bar q$\, QED mesons.

In order to infer the importance of the second quark condensate term
relative to the massless quark limit arising from the confining
interaction in (\ref{qcd}), we tabulate in Table I the results of the
hadron mass values obtained in the massless quark limit.  We observe
that for the pion mass, the massless quark limit is zero, and the pion
mass arises only from the second quark condensate term.  The
importance of the quark condensate diminishes as the hadron masses
increases to $\eta$ and $\eta'$.  Thus, in experiments in which the
quark condensate may be affected by the environment in which the
hadrons are produced, as for example in a hot quark gluon plasma at
various temperatures, the degree of quark condensation may decrease as
the temperature increases, resulting in a downward shift of the hadron
mass towards the massless quark limit.  The massless quark limit may
be reached at a critical temperature when the chiral symmetry is fully
restored with the absence of a quark condensate.  As a consequence,
the shift in the hadron masses may be a signature of the restoration
of chiral symmetry.  It is interesting to note in the QCD meson case
that the absence of the quark condensate leads to the pions acquiring
a zero mass, which is an indication of the dissociation of the pion
into a massless quark and antiquark pair when chiral symmetry is
restored.  It is reasonable to consider this to be an indication that
the occurrence of chiral symmetry restoration for a pion occur
alongside with the occurrence of the deconfinement of the quark and
the antiquark in the pion.

\subsection{Open string description of  the QED mesons \label{sec25}}

Having confirmed the approximate validity of the open string
description of QCD mesons in 1+1 dimensions, we proceed to extrapolate
to the unknown region of the open string $q\bar q$ QED mesons.  In
this subsection (\ref{sec25}), we restrict ourselves to the QED
interaction with the interaction label superscript set implicitly to
$\lambda$=0 for QED.  The mass scale in (\ref{14}) gives $m^{\qed}\!\!\!\!
\sim 47 {\rm ~MeV}\gg m_u, m_d$, but $m^{\qed}$ is comparable to
$m_s$.  In the treatment of QED mesons, it is only necessary to
include $u$ and $d$ quarks and antiquarks, with $N_f =2$.
 
We denote flavor states $\phi_1=|u \bar u \rangle$, $\phi_2=|d \bar d
\rangle$, and construct the physical isoscalar $|\Phi_1^{\qed}\rangle$
and the isovector $|\Phi_2^{\qed}\rangle$ states as
\begin{eqnarray}
|\text{(isoscalar)}I=0,I_3=0\rangle&=\Phi_1=({\phi_1+\phi_2})/{\sqrt{2}},
\nonumber\\ |\text{(isovector)}I=1,I_3=0\rangle&=\Phi_2=({\phi_1-\phi_2})/{\sqrt{2}}.
\end{eqnarray}
They are related by $\Phi_i$=$\sum_f D_{if}\phi_f$ and
$\phi_i$=$\sum_f D_{if}\Phi_i$,
\begin{eqnarray}
\begin{pmatrix}
 \Phi_1\\ \Phi_2
\end{pmatrix}
\!\!=\!\!\begin{pmatrix} \frac{1}{\sqrt{2} } & +\frac{1}{\sqrt{2} }
\\ \frac{1}{\sqrt{2} } & - \frac{1}{\sqrt{2} }
\end{pmatrix}
\!\!
\begin{pmatrix}
 \phi_1\\ \phi_2
\end{pmatrix}
, ~~~~~
\begin{pmatrix}
 \phi_1\\ \phi_2
\end{pmatrix}
\!\!=\!\!\begin{pmatrix} \frac{1}{\sqrt{2} } & + \frac{1}{\sqrt{2} }
\\ \frac{1}{\sqrt{2} } & - \frac{1}{\sqrt{2} }
\end{pmatrix}
\!\!
\begin{pmatrix}
 \phi_1\\ \phi_2
\end{pmatrix}.
\end{eqnarray}
The mass formula (\ref{230}) for the mass of $\Phi_i$ becomes
\begin{eqnarray}
m_I^2&=\left [ \frac{Q_u+(-1)^IQ_d}{\sqrt{2}} \right ]^2
\frac{4\alpha_{{}_{\rm QED}}}{ \pi R_T^2} + m_{ud} e^\gamma \mu^\qed,
\label{240}
\end{eqnarray}
where the mass scale
$\mu^\qed$ for QED mesons is not known.  From the results for QCD mesons in
(\ref{231})-(\ref{233}), we expect an analogous relationship relating
the mass scale $\mu^\qed$ and the quark condensate for QED mesons,
 \begin{eqnarray}
e^\gamma \mu^\qed \propto |\langle 0|\bar q q|0 \rangle|_{_{\qed}},
\end{eqnarray}
where $|\langle 0|\bar q q|0 \rangle|_{_{\qed}}$ is the quark
condensate in the presence of QED gauge interactions between the quark
and the antiquark.  At the present stage of our development, there is
no experimental information to quantify $|\langle 0|\bar q q|0
\rangle|_{\qed}$.  It is also not known whether such a quark
condensate term may be affected by the environment such as the
temperature or the mechanism of QED meson production.  It is only
known that for QCD mesons, such a quark condensate term is related to
the pion mass.  We expect that the quark condensate term should depend
on the coupling constant $g_{\2d}$ of the gauge interaction,
whether it be QCD or QED.  We note that the first term in (\ref{230})
depends on the coupling constant as $(g_{\2d})^2$, and in our
comparison in QCD, the quark condensate term is just as important as
the massless quark limit for the lightest QCD meson.  Thus, pending
future amendments, we assume that the second quark condensate term  in (\ref{240}) for QED is of
order $(g_{\2d}^{\qed})^2$, the same as the first massless quark limit
term.  In this case, we have
\begin{eqnarray}
 \frac{ (\mu^\qed  \text{ in the quark condensate term  for QED )}}
 {(\mu^\qcd \text{ in the quark
     condensate term  for QCD)}}
 &=&\frac{(g_{\2d}^{\qed})^2}{(g_{\2d}^{\qcd})^2}=\frac{\alpha_{{}_{\rm QED}}}{\alpha_s}.
\label{242}
\end{eqnarray}
From Eqs.\ (\ref{231}), (\ref{240}), and (\ref{242}), it is then
reasonable to consider the phenomenological semi-empirical mass
formula for QED mesons as
\begin{eqnarray}
m_I^2&=\left [ \frac{Q_u+(-1)^IQ_d}{\sqrt{2}} \right ]^2
\frac{4\alpha_{{}_{\rm QED}}}{ \pi R_T^2} + m_\pi^2\frac{\alpha_{{}_{\rm QED}}}{\alpha_s} .
\label{qed}
\end{eqnarray}
Here, the first term is the massless quark limit  arising from the
confining interaction between the quark and the antiquark, with 
 $Q_u$=2/3, $Q_d$=$-$1/3, and $\alpha_{{}_{\rm QED}}=$1/137.
It depends only on the flux tube radius $R_T$.
The second term arises from the quark masses  and  the quark
condensate in the presence of the QED interaction.

In applying the above mass formula for QED2 mesons, we extrapolate
from the QCD sector to the QED sector by using those $R_T$ and
$\alpha_s$ parameters that describe well the $\pi^0$, $\eta$ and
$\eta'$ QCD mesons in Section \ref{sec24}.  We list the theoretical
masses of the neutral, $I_3$=0, $S$=0 QED mesons obtained by
Eq.\ (\ref{qed}) in Table I.  We find an $I$=0 isoscalar QED meson at
$m_{\rm isoscalar}^{\qed}$=17.9$\pm$1.5 MeV and an $(I$=1,$I_3$=0)
isovector QED meson at $m_{\rm isovector}^{\qed}$=36.4$\pm$3.8 MeV.
As the $I^G(J^{PC})$ quantum numbers of the QCD mesons are known, we
can infer the quantum numbers of the corresponding QED mesons with the
same $I$ and $S$ by analogy. Such an inference by analogy provides a
useful tool to determine the quantum numbers and some electromagnetic
decay properties of QED mesons. Using such a tool, we find that the
isoscalar QED meson has quantum numbers $I^G(J^{PC})$=$1^-(0^{-+})$
and the isovector $I_3$=0 QED meson has quantum numbers
$I^G(J^{PC})$=$0^+(0^{-+})$.  Within the theoretical and experimental
uncertainties, the matching of the $I(J^\pi)$ quantum numbers and the mass
 may make the isoscalar
QED meson a good candidate for the X17 particle emitted in the decay
of the $0(0^-)$ state of He$^4$ in \cite{Kra19}.

It is interesting to note that for the $I_3$=0, $S$=0 QCD and QED
mesons, the mass ordering for the isovector and isoscalar QCD mesons
is the reverse of that for the isovector and isoscalar QED mesons.
This arises because in QCD mesons there is no difference in the
magnitudes and the signs of the color charges of the $u$ and $d$
quarks, with $Q_u^\qcd$=$Q_d^\qcd $=1, whereas in QED mesons there is
a difference in the magnitudes and the signs of the electric charges
of the $u$ and $d$ quarks, with $Q_u^\qed$=2/3 and $Q_d^\qed$=$-$1/3.

In order to show the effects of the second quark condensate term
relative to the massless quark climit in (\ref{qed}), we
tabulate in Table I the results of the QED meson masses values
obtained in the massless quark  limit.  One observes that the
mass of the isoscalar QED meson with the quark condensate is
17.9$\pm$1.5 MeV but it is reduced to 11.2$\pm$1.3 MeV in the massless
quark limit without the quark condensate.  The QED meson masses may be
affected by the environment in which the mesons are produced, as for
example in a hot quark gluon plasma at various temperatures.  The
degree of QED quark condensation is expected to decrease as the
temperature increases, resulting in a downward shift of the meson
mass towards the massless quark limit as the temperature increases.
The massless quark limit may be reached at a critical temperature when
the spontaneously broken chiral symmetry is fully restored with the
absence of a quark condensate.  When that happens, the mass of the
isoscalar QED meson may be shifted to a lower value of about
11.2$\pm$1.3 MeV.  As a consequence, the shift in the hadron masses
may be a signature of the restoration of chiral symmetry.  The quark
condensate may also be affected by the mechanism for the production of
the QED mesons.  If an isoscalar QED meson is produced in a mode
without a quark condensate, then its mass may be shifted downward to
11.2$\pm$1.3 MeV.  It is interesting to note that in the QED meson
case the absence of the quark condensate leads to a decrease of the
isoscalar QED meson mass from 17.9$\pm$1.5 MeV to 11.2$\pm$1.3 MeV,
but the mass remains non-zero at 11.2$\pm$1.3 MeV, indicating that the
QED confinement between the quark and the antiquark remains
operational even when chiral symmetry is restored.  It is reasonable
to consider then the possibility that the occurrence of chiral
symmetry restoration and the occurrence of the deconfinement of an
isoscalar QED meson take place at different temperatures of the
environment.

In Table I, we have also listed the hypothetical X17 particle observed
in \cite{Kra16,Kra19} and the hypothetical E38 particle observed in
\cite{Abr12,Abr19} as possible QED meson candidates because their
measured invariant masses appear to be close to the masses of the
predicted isoscalar and isovector QED mesons, respectively.  It will
be of great interest to confirm or refute the existence these
hypothetical particles by independent experimental investigations, as
a test of the QED meson concepts.

\section{Production of the QCD and QED mesons}

How are QCD and QED mesons produced?  We can consider the interaction
of particles $A$ and $B$ in Fig.\ \ref{fig1}, where $A$ or $B$ can be
a nucleon in free space, a nucleon inside a nucleus, or a meson.  The
double solid lines in Fig.\ \ref{fig1} represent a diquark in the case
of a nucleon, or an antiquark in the case of a meson.  The production
of QCD and QED mesons $C_i$ and $D_i$ by the interaction of particles
$A$ and $B$ may be described by Feynman Diagrams \ref{fig1}(a),
\ref{fig1}(b), and \ref{fig1}(c).  Diagram \ref{fig1}(a) is for low
energies, Diagram \ref{fig1}(b) is for intermediate energies, and
Diagram \ref{fig1}(c) is for high energies.  The produced particles
may be either QCD or QED mesons if the available energy is above the
QCD meson mass threshold.  On the other hand, if the available energy
is below the pion mass threshold, as in the decay of the excited $^4$He
and $^8$Be in \cite{Kra16,Kra19}, the produced particle $C_1$ 
in  Diagram
\ref{fig1}(a)
can only be a QED meson.
Diagrams \ref{fig1}(b) and \ref{fig1}(c) may describe the simultaneous
production of QCD mesons and QED mesons in the anomalous soft photon
phenomenon in high-energy hadron-proton reactions, when a fraction of
the produced particles $C_i$ and $D_i$ are QED mesons while the
dominant fraction are QCD mesons.  The produced QED mesons
subsequently decay into $e^+ e^-$ pairs as anomalous soft photons in
\cite{Bot91,Ban93,Bel97,Bel02} or as $\gamma \gamma$ pairs in \cite{Abr12,Abr19,Ber11,Ber14,Sch11,Sch12}.

\begin{figure} [H]
\centering
\vspace*{-0.0cm}\hspace*{-0.3cm}
\includegraphics[angle=0,scale=0.80]{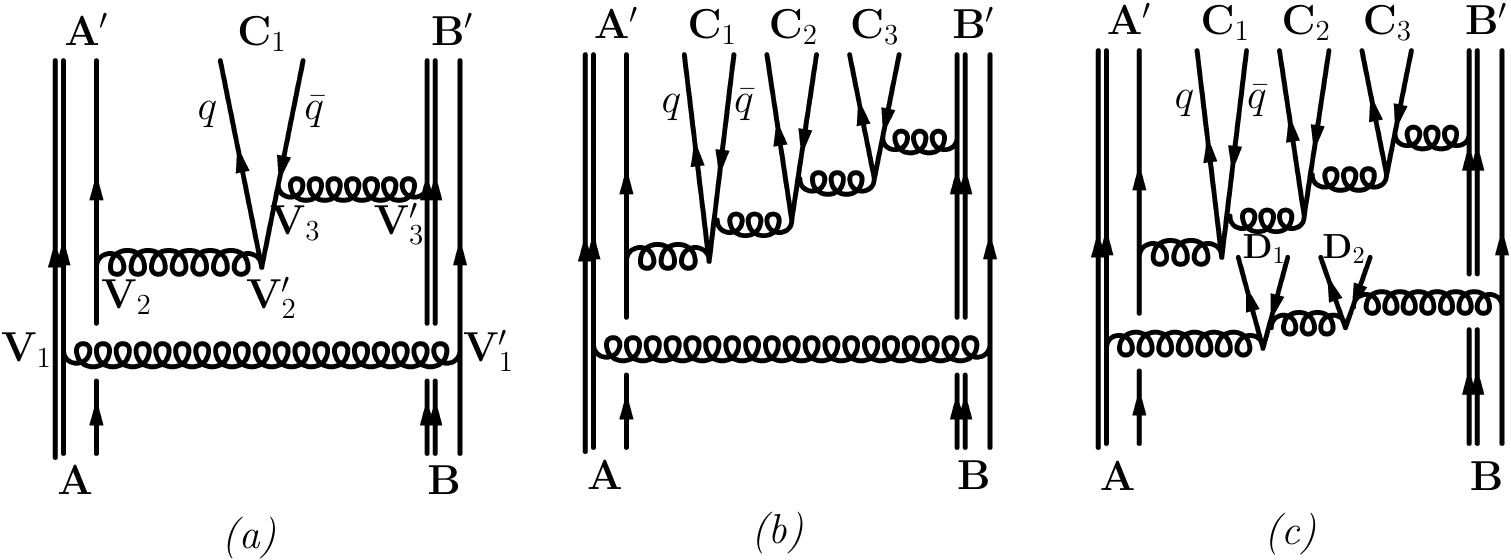} 
\vspace*{-0.3cm}
\caption{Feynman diagrams for the production of QCD and QED mesons :
  ({\it a})$ A+B \to A'+B'+C_1$ at low energies, ({\it b})$ A+B \to
  A'+B'+C_1+C_2+C_3...$ at intermediate energies, and ({\it c})$ A+B
  \to A'+B'+C_1+C_2+C_3 +...+D_1+D_2+...$ at high energies.  The
  double lines represent a diquark in the case of  a baryon and an antiquark in the case of  a
  meson.  }
\label{fig1}
\end{figure}

To describe the decay of the excited $0(0^-)$ state of $^4$He at an
excitation energy of 21.02 MeV, we can consider the preparation of the
excited state of $^4$He by pulling a proton out of a tightly bound
$^4$He nucleus, in the configuration of a stretched string, with the
proton and the remainder $^3$H$^*$ at the two ends of the string.  We
can represent the outside proton as the particle $A$ in Diagram
\ref{fig1}(a), and a nucleon in the much heavier $^3$H$^*$ core
remainder as the particle $B$ in Diagram \ref{fig1}(a).  The strong
binding (of 21.01 MeV) between the proton and the core remainder leads
to a strong interaction that polarizes the spatial region between them
in the stretched string, leading to the vacuum polarization and the
creation of a $q \bar q$ pairs at the vertex $V_2'$ in Diagram
\ref{fig1}(a).  The creation of the quark and antiquark pair is
energetically possible because of the masses of the up and down quarks
are only of order a few MeV \cite{PDG19}.  The available energy of
21.01 MeV is above the $q\bar q$ pair production threshold but below
the pion mass threshold.  The produced $q\bar q$ pair can only
interact with QED interactions to form a QED meson at the appropriate
energy.  By way of such pair production in vacuum polarization, there
may be the occasional production of a quantized open string isoscalar
$q\bar q$ \,QED meson $C_1$ with the proper quantum numbers at the
appropriate QED meson energies.  Similarly we can envisage the
possible production of X17 from the 18.15 MeV 0(1$^+$) excited state
of $^8$Be that is prepared by stretching out a proton from one of the
two $^4$He nuclei in the $^8$Be nucleus.  Vacuum polarization with the
production of a $q \bar q$ pairs may lead to the production of the
open string isoscalar 0($0^-)$ QED meson, coming out in the $l$=1
partial wave.

All of the initial and final states of the reaction $(AB) \to (A'B') +
C_1$ in Diagram \ref{fig1}(a) involve colorless particles, and it is
instructive to follow the color flow of the intermediate states to see
how such a reaction proceeds from colorless initial particles $A$ and
$ B$ to colorless final particles $A'$, $B'$ and $C_1$ in Diagram
\ref{fig1}(a).  After the emission of the gluon at the point $V_1$
from the diquark of nucleon $A$, the intermediate state of $A$
acquires a color, which is however bleached to become the colorless
$A'$ upon the emission of the gluon at $V_2$.  Similarly, the nucleon
$B$ becomes colored upon absorbing a gluon at vertex $V_1'$, but the
color is bleached upon the absorption of a gluon at $V_3'$ to lead to
the final colorless nucleon $B'$.  The produced $q\bar q$ pair is
initially colored because it originates from the gluon at vertex
$V_2'$.  The color of the $q\bar q$ pair is bleached to become the
colorless final state $C_1$ upon the emission of a gluon at $V_3$.
The important ingredient for the occurrence of such a reaction process
is the availability of the excitation energy and the small values of
the up and down quark masses which facilitate the production of the
$q\bar q $ pair.
 
Recently, a boson particle with invariant a mass of about 38 MeV
(labeled as the hypothetical E38 particle) decaying into $\gamma
\gamma$ pairs has been observed at Dubna in high-energy $\{ p{\rm C},
d{\rm C}, d{\rm Cu} \}$$ \to$$ \gamma \gamma X$ collisions at proton
and deuteron incident energies of 5.5, 2.76, and 3.83 GeV per nucleon,
respectively \cite{Abr12,Abr19}.  The observed E38 mass coincides,
within the experimental and theoretical uncertainties, with the
predicted mass of 36.4$\pm$3.8 MeV for the isovector
$[I(J^\pi)=0(0^-),I_3=0]$ QED meson, making the isovector QED meson a
good candidate for the E38 particle.  Diagrams \ref{fig1}(b) or
\ref{fig1}(c), in which one of the final particles is the isovector
QED meson, may describe the production of the E38 particle in these
high-energy reactions.

In another set of experiments, the raw data of the $\gamma \gamma$
invariant mass spectra in exclusive measurements in $pp$$\to$$p (
\pi^+ \pi^- \gamma \gamma) p$ \cite{Ber11,Ber14} and $\pi^-p$$\to$$\pi^- (
\pi^+ \pi^- \gamma \gamma) p$ reactions at $p_{\rm lab}$=193 GeV/c at
COMPASS \cite{Sch11,Sch12} exhibit structures at around 10-15 MeV and
38 MeV.  It has been argued that these structures may arise from
experimental artifacts \cite{Ber12}.  On the other hand, it has also
been suggested that even after the background data are subtracted,
there may remain a significant structure at 38 MeV \cite{Bev12}.  What
may be of great interest are the following experimental facts: (i) the
locations of the energies of these structures at 10-15 MeV and 38 MeV
fall within the vicinity of the predicted masses of the isoscalar QED
meson and isovector QED meson respectively, (ii) the peak at around
10-15 MeV in the $\gamma \gamma$ invariant mass spectrum in
$pp$$\to$$p ( \pi^+ \pi^- \gamma\gamma) p$ \cite{Ber11,Ber14} appears
to be substantially above the neighboring background distribution
\cite{Ber11,Ber14}, (iii) diphoton resonance at 38 MeV has been
observed high-energy $\{ p{\rm C}, d{\rm C}, d{\rm Cu} \}$$ \to$$
\gamma \gamma X$ reactions \cite{Abr12,Abr19}, and (iv) the occurrence
of the structure appears in both $pp$ and $\pi^- p$ collisions,
indicating that the phenomenon may be more general than a single
projectile-target combination.  If one looks at the $pp$ and $\pi^- p$
reactions at COMPASS from the theoretical viewpoint of Diagram
\ref{fig1}(b), it is possible that while most of the produced $q\bar
q$ pairs in the set of $(C_1, C_2, C_3)$ are QCD mesons, it cannot be
excluded that some of the produced $C_i$ particles among the set of
three $C_i$ particles may be an isoscalar or isovector QED meson with
a mass of order 17 MeV or 38 MeV.  Whether these structures at 10-15
MeV and 38 MeV represent genuine particle states remains a subject for
further studies.  Future investigations in the region of low $\gamma
\gamma$ invariant masses will provide additional tests to confirm or
refute the proposed concept of the open string QED mesons.

\section{Transverse momentum distribution of anomalous soft photons }

We would like to examine how  the production of quantized QED mesons may
be consistent with the anomalous soft photon phenomenon.
In Fig.\ \ref{fig2},
the solid circular points give the experimental $dN/dp_T$ data of
soft photons  measured 
as $e^+ e^-$ pairs  after subtracting the experimental background.
Fig.\ \ref{fig2}(a) is  the WA102 results for
 $pp$ collisions at $p_{\rm
  lab}$=450 GeV/c  from  Belogianni $et~al.$ \cite{Bel02}, and
Fig.\ \ref{fig2}(b) is the DELPHI results  for $e^+e^-$ annihilation at the $Z_0$
  mass of 91.18 GeV \cite{DEL06}. 
They are 
 exclusive measurements in which
the momenta of all participating charged particles are measured.  
The knowledge of the momenta of 
all initial and final  charged particles allows an accurate determination of the bremsstrahlung $dN/dp_T$ distributions shown as triangular points in Fig.\ (\ref{fig2}). 
As shown in Fig.\  \ref{fig2},
the observed yields of  soft photons  exceed 
the bremsstrhlung contributions substantially in both $pp$ collisions in
 \ref{fig2}(a)   
and $e^+e^-$ annihilations  in \ref{fig2}(b).   The  excess  soft photons as $e^+e^-$ pairs 
  constitute the soft photon anomaly
that is the subject of our attention.

\begin{figure} [h]
\centering
\includegraphics[angle=0,scale=0.90]{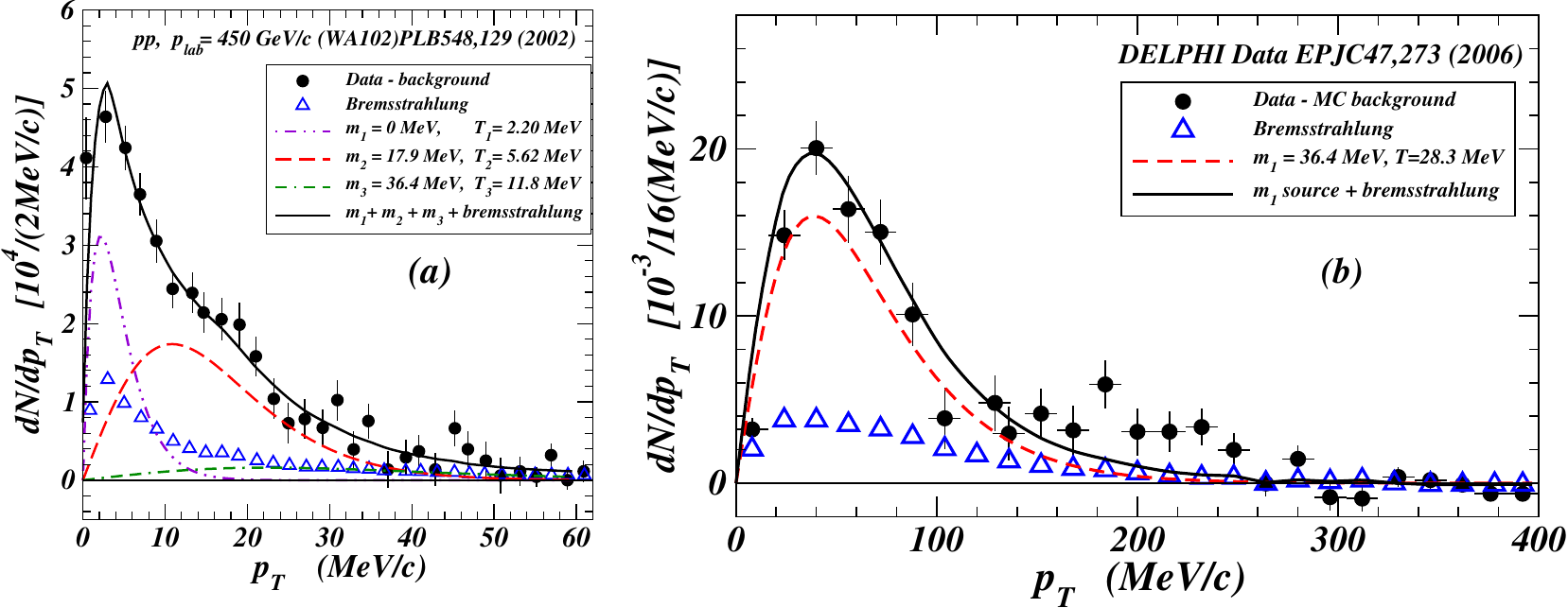}
\caption{ (a) Anomalous soft photon $dN/dp_T$ data from $pp$
  collisions at $p_{\rm lab}$=450 GeV/c obtained by Belogianni
  $et~al.$ \cite{Bel02}.  (b) Anomalous soft photon $dN/dp_T$ data
  from the DELPHI Collaboration for $e^+e^-$ annihilation at the $Z_0$
  mass of 91.18 GeV \cite{DEL06}.  The solid circular points represent
  the experimental data after subtracting the experimental background,
  and triangle points represent the deduced bremsstrahlung
  contributions.  The total theoretical yields in the thermal model from produced bosons and  the additional bremsstrahlung contributions are shown as
  solid curves.  The component yields from different masses of the
  thermal model are shown as separate curves.}
\label{fig2}
\end{figure}

We would like to inquire how the production of the QED mesons
may be consistent with the excess $e^+e^-$ pairs in the anomalous soft photon phenomenon.
For such an investigation, we rely on the thermal model
which describes well the transverse momentum distributions 
in the production of hadrons of different masses 
 in high-energy $pp$ collisions \cite{Hag65,Abe07,Abe09,Ada11}.
We shall assume that the validity of the thermal model can be extended
from the production of QCD mesons to the production of QED mesons
whose decay products are assumed to appear as anomalous soft photons.
In such a thermal model, the transverse momentum distribution of the
produced QED mesons is related to the produced QED meson mass $m$ by \cite{Hag65}
\begin{eqnarray}
\frac{dN}{ p_T dp_T}=A e^{-\sqrt{m^2+p_T^2}/T}.
\end{eqnarray}
The 
contribution    to the total $dN/dp_T$ from each boson of mass $m$  is proportional to $p_{{}_T}e^{-\sqrt{m^2+p_T^2}/T}$ which    is zero at $p_T$=0 and has a peak  at the location $p_T$ given by 
\begin{eqnarray}
p_T^2 = \frac{1}{2} [ T^2 + \sqrt{T^4 +4T^2m^2}].
\end{eqnarray}
 If $m=0$, then $dN/dp_T$ peaks at $p_T=T$.  If $m$ is much greater
 than $T$, then $dN/dp_T$ peaks at $p_T \sim$ $\sqrt{mT}$.  Hence, for each
 contributing boson mass component, the thermal model gives a distribution that starts at zero at $p_T$=0 and reaches a peak of $dN/dp_T$ and decreases from the peak.
 The total $dN/dp_T$ is a sum of contributions from different QED
 mesons and boson components,
\begin{eqnarray}
\frac{dN}{p_T dp_T}\!=\!  \sum_{i}A_i e^{-{\sqrt{m_i^2+p_T^2}/T_i}}.
\end{eqnarray}
There can be as many contributing bosons as the number of underlying peaks in the $dN/dp_T$ spectrum. 
While different decompositions of
the spectrum into different masses (and peaks) are possible, the structure of the $dN/dp_T$ data appears
to require many components in Fig.\ 1(a) and only a single component in
Fig. \ref{fig2}(b).    In the thermal model
analysis of the $pp$ data  in Fig. \ref{fig2}(a), 
we note that there appears to be a boson component of real
photons with $m_1$=0. 
 Because of the $m_i T_i$ ambiguity\footnote{In
  fitting the thermal model, the $m_iT_i$ ambiguity gives different
  values of $m_i$ for different values of $T_i$ without changing
  significantly the overall quality of the fitting.} associated with
the product of the meson mass $m_i$ and the temperature $T_i$, we are
content with only a consistency analysis.  
We assume that QED isoscalar and isovector mesons with masses as given by Table I are produced in the collision, and their subsequent decay into $e^+ e^-$ pairs give rise to the excess  $e^+ e^-$ pairs observed as
anomalous soft photons.
Allowing other parameters to vary,  the thermal model fit in Fig.\ \ref{fig2}(a) is
obtained with parameters $A_1$=$3.85\times10^4$$/(2{\rm
  MeV}/c$), $T_1$=2.20 MeV, $A_2$=6.65$\times10^4$$/(2({\rm MeV}/c$),
$T_2$=5.62 MeV, $A_3$=0.266$\times10^4$$/(2({\rm MeV}/c$), and
$T_3$=11.8 MeV, where the different components also shown as separate
curves.    
Adding the contributions from the three components onto the
bremsstrahlung contributions yields the total $dN/dp_T$  shown as the solid curve.
The comparison in Fig.\ \ref{fig2}(a)  indicates that  the $pp$ 
data are
consistent with a photon component and a boson component with a mass around 17 MeV.  The  magnitude of the $m_3$=36.4 MeV  component is of the same order as the bremsstrahlung contribution or the noise level, and is rather uncertain.  
 In Fig.\ \ref{fig2}(b),
the addition of the 
single component with $A_1$=2.7$\times$10$^{-3}$$/(16{\rm MeV}$/c),
$m_1$=36.4 MeV, and $T_1$=28.3 MeV onto the bremsstrahlung
contributions gives a consistent description of the soft photon data in $e^+e^-$ annihilations as shown as the solid curve.

The component with $m_1$=0 in Fig.\ \ref{fig2}(a) may be associated
with the decay of the QED mesons into two photons.  If so, it will be
of interest to measure the $\gamma \gamma$ invariant mass to look for
diphoton resonances, as carried out in
\cite{Abr12,Abr19,Ber11,Ber14,Sch11,Sch12}.  The $m_2$=17.9 MeV
components in Fig.\ \ref{fig2}(a) and the $m_3$=36.4 MeV component in
Fig.\ \ref{fig2}(b) may be associated with the predicted isoscalar and
isovector QED mesons of Table \ref{tb1}.  If so, a measurement on the
invariant masses of the $m_2$ and $m_3$ components will be of great
interest to confirm the existence of these QED mesons.  The recent
reports of the observation of a hypothetical E38 boson at 38 MeV and
the structures in the $\gamma \gamma$ invariant masses at 10-15 MeV
and 38 MeV \cite{Abr12,Abr19,Ber11,Ber14,Sch11,Sch12} provide
encouraging impetus for further studies.

We can envisage how QED mesons may be produced alongside with QCD
mesons in the soft photon phenomenon in high-energy particle
collisions
\cite{Chl84,Bot91,Bot91,Ban93,Bel97,Bel02,DEL06,DEL08,Per09,DEL10} in
which meson production has been well understood as a string
fragmentation process \cite{Bjo73,Cas74,Art74,And83}.  Many $q\bar q$
strings may be produced when the string joining a valence quark and a
valence antiquark are pulled apart.  
Because quarks and
antiquarks interact with both QCD and QED interactions and they may
form meson states, both QCD mesons and QED mesons may be
simultaneously produced during the high-energy string fragmentation
process. 
 Such a  
simultaneous production of 
 QCD mesons and QED mesons
 can be  alternatively described by the production mechanism of
Diagrams \ref{fig1}(b) and \ref{fig1}(c), in which the final produced particles
$C_i$ or $D_i$ may be QCD or QED mesons.  
 The produced QED mesons
subsequently decay into $e^+e^-$ and $\gamma\gamma$ pairs which may
appear as excess $e^+e^-$ and $\gamma \gamma$ yields to accompany the
produced QCD mesons \cite{Per09,DEL10,Won10}.  The QED mesons and
their decay $e^+e^-$ products will not be produced when hadrons are
not produced in $e^+$+ $e^-$$\to$ $\mu^+$+ $\mu^-$ bremsstrahlung
\cite{DEL08}.

There remain many unresolved questions and uncertainties regarding the
anomalous soft photons as presented in Fig. 2.  The thermal model
analysis only provides an approximate hint on the possible
contributions from many mass components, as the fitting of the boson
masses in the thermal model contains ambiguities associated the
determination of the masses and temperatures.  It is also not known
why the two components of $m_1$=0 and $m_1$=16.9 MeV in the $pp$
measurement in Fig 1(a) are not present in $e^+ e^-$ annihilation
DELPHI measurement in Fig.\ 1(b).  What may be important however are
the cross section enhancements in many transverse momentum regions,
suggesting possible masses where quantized boson masses may occur.
The additional measurements of the invariant masses of the excess
$e^+e^-$ and $\gamma \gamma$ pairs of the decay photon energies in the
neighborhood of these enhancements will reveal whether or not the
excess $e^+e^-$ and $\gamma \gamma$ pairs represent quantized bosons,
to test the concepts of the QED mesons.

\section{Behavior of a massive QED meson assembly } 

The QCD and QED mesons with massless quarks in 1+1 dimensions cannot
decay as the quark and the antiquark execute yo-yo motion along the
string.  As the string is an idealization of a flux tube, the
structure of the flux tube must be taken into account in the physical
processes in 3+1 dimensions.  The quark and the antiquark at different
transverse coordinates in the tube traveling from opposing
longitudinal directions can make a sharp change of their trajectories
turning to the transverse direction where the quark and the antiquark
can meet and annihilate, leading to the emission of two photons at the
vertices $V_1$ and $V_2$ as depicted in Fig.\ \ref{fig3}(a).  The
coupling of these photons to an electron pair as shown in
Fig.\ \ref{fig3}(b) leads further to the decay of the QED meson into
an electron-positron pair.  Thus, by the consideration of the
transverse structure of the flux tube, the QED mesons can decay into
photons and electron-positron pairs in 3+1 dimensions.

\begin{figure} [h]
\centering \includegraphics[angle=0,scale=0.70]{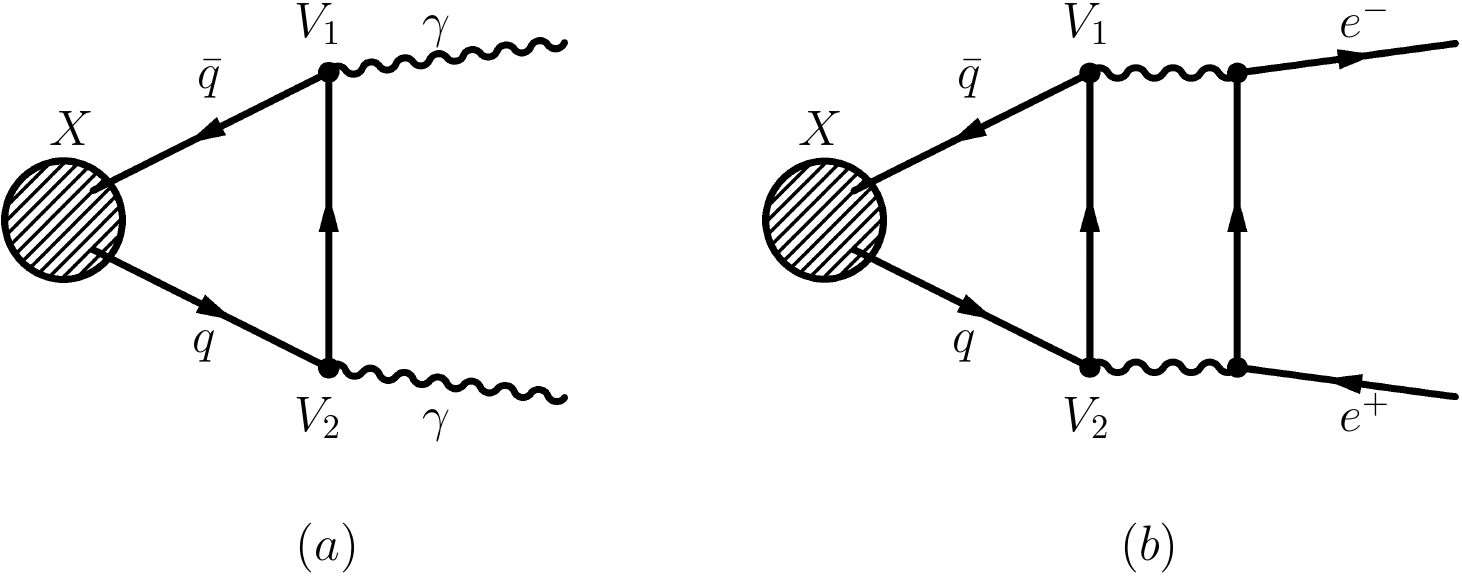}
\caption{Decay of the QED meson X into ({\it a}) a $\gamma \gamma$
  pair, and ({\it b}) an $e^+ e^-$ pair.  }
\label{fig3}
\end{figure}

An astrophysical object consisting of a large assembly of isoscalar
$0(0^-)$ QED mesons such as the X17 particle with a mass $m_X$=17 MeV
will be an electron-positron and gamma-ray emitter.  If the
temperature of such an assembly is low, it can form a Bose-Einstein
condensate.  The mode of emission, the emission energies, and the lifetimes  depend on
the gravitational energy of the assembly.  Such assemblies of QED
mesons present themselves as good candidates as $e^+$$e^-$ emitters,
gamma-ray emitters, or the primordial cold dark matter.  We would like to
make estimates on the constraints on masses and radii of such
assemblies where they may be found.

We consider an assembly of $A$ number of $m_X$ QED mesons of mass
$M_A$$\equiv$$M$ and we place a test QED $m_X$ meson at the surface of
the assembly at radius $R$, the mass $M_{A+1}$ of the combined system
is
\begin{eqnarray}
M_{A+1} = M_A + m_X - \frac{G M_A m_X}{Rc^2},
\end{eqnarray}
where $G$ is the gravitational constant.  The $Q$ value for the test
QED meson at the surface of the ($A$+1) assembly to decay into an
electron-positron pair is
\begin{eqnarray}
\hspace*{-0.7cm} Q( (A+1) \!\to\! A+e^+e^-)=\! m_X c^2\! -\!\frac{G
  M_{A} m_X}{R}\! -\! 2m_e c^2, ~
\end{eqnarray}
and the $Q$ value for the test QED meson to decay into two photons is
\begin{eqnarray}
Q((A+1) \to A +2\gamma)= m_X c^2 -\frac{G M_{A} m_X}{R}.
\end{eqnarray}
Thus, the QED meson $m_X$ will not decay into an electron-positron
pair when the mass and radius of the assembly satisfy
\begin{eqnarray}
\frac{M }{R}> \frac{c^2}{G}\left (1-\frac{2m_e }{m_X} \right ),
\label{MM0}
\end{eqnarray}
and the QED meson $m_X$ will not decay into two photons nor an
$e^+e^-$ pair when $M$ and $R$ satisfy
\begin{eqnarray}
\frac{M }{R}> \frac{c^2}{G}.
\label{MM1}
\end{eqnarray}
Upon using the mass and the radius of the sun as units, it is
convenient to define a dimensionless boundary value $B_0$ given by
\begin{eqnarray}
B_0=\frac{ c^2R_\odot }{ G M_\odot}=4.71\times 10^{5}.
\end{eqnarray}
A QED meson assembly will behave differently depending on its $M/R$
values as follows:
\begin{enumerate}

\item
The QED meson assembly will emit electron-positron pairs and gamma
rays if
\begin{eqnarray}
  B_0 \left (1-\frac{2m_e }{m_X} \right ) > \frac{M /M_\odot
  }{R/R_\odot }.
\end{eqnarray}

\item
The QED meson assembly will emit only gamma rays but no $e^+$$e^-$
pairs, if
\begin{eqnarray}
B_0 > \frac{M /M_\odot }{R/R_\odot } > B_0 \left (1-\frac{2m_e }{m_X}
\right ).
\end{eqnarray}

\item
The QED meson assembly will not emit $e^+$$e^-$ pairs nor gamma rays,
if
\begin{eqnarray}
 \frac{M /M_\odot }{R/R_\odot } > B_0 ,
\label{MM4}
\end{eqnarray}
which is essentially the condition for a QED meson black hole.

\end{enumerate}
The above boundaries characterize the properties of QED meson
assemblies as $e^+e^-$ emitters, gamma-ray emitters, or dark matter.
An assembly of QED mesons satisfying (\ref{MM1}) (which is the same as
or (\ref{MM4})) can be a good candidate for a primordial cold
blackhole dark matter, as it is non-baryonic, created at the
hadronization stage of the quark-gluon plasma phase transition, and
not from a stellar collapse.

\section{Conclusions and discussions}

Many interesting questions have been brought into focus with the
observations of the X17 particle \cite{Kra16,Kra19}, the E38 particle
\cite{Abr12,Abr19}, and the related anomalous soft photon phenomenon
\cite{Chl84,Bot91,Ban93,Bel97,Bel02,DEL06,DEL08,Per09,DEL10}.  Our
investigations provide answers to some of these question by linking them together in a coherent framework of open string QED mesons.
However, many
unanswered questions remain and these questions  will  require further theoretical and
experimental studies.  The central questions are (i) whether quarks
and antiquarks interacting with the QED interaction can form confined
and bound QED mesons in the mass range of many tens of MeV as
suggested in \cite{Won10}, (ii) whether the X17 and E38
particles, and the parent particles of anomalous soft photons are QED mesons,
and (iii) whether there can be additional tests to confirm or refute
the existence of the open string QED mesons.

On the theoretical side, it can
be argued that Schwinger already showed that massless fermions and
antifermions interacting with QED interactions in 1+1 dimensions can
form confined and bound boson states \cite{Sch62,Sch63}.  As a quark
and an antiquark cannot be isolated and the intrinsic motion of 
a $q \bar q$ system in its 
lowest-energy  states lies predominantly in 1+1 dimensions  \cite{Ven68}-\cite{Cas74}, an open string description in
1+1 dimensions can be applied to study the 
$q \bar q$ system in QCD and QED interactions.  Such interactions lead
to confined and bound physical QCD and QED meson states
\cite{Won10,Won11,Won14}.  We show that the $\pi^0$, $\eta$, and
$\eta'$ can be adequately described as open string QCD mesons in 1+1
dimensions, when we properly take into account the relevant physical
effects.  The extrapolation from the QCD mesons 
to the QED mesons with  $q$ and $\bar q$ interacting with QED interactions lead to an isoscalar
$I(J^\pi)$=0(0$^-)$ QED meson state at 17.9$\pm$1.5 MeV and an
isovector $(I(J^\pi)$=1($0^-), I_3$=0) QED meson states at
36.4$\pm$3.8 MeV.

On the experimental side,
possible occurrence of the QED mesons 
can be confirmed or refuted by searching for
the decay product of  $e^+ e^-$ and $\gamma \gamma$ pairs.  The decay
experiments carried out in \cite{Kra16,Kra19} are good examples for
such explorations with the observation of 
the  X17 particle at  an 
$e^+e^-$ 
invariant mass of 17 MeV, 
from the  
the decay of the excited 0(0$^-)$ state of He$^4$.
The
matching of the quantum numbers and the masses, within the
experimental and theoretical uncertainties, make the isoscalar
0(0$^-)$ QED meson a good candidate for 
the X17 particle.
There is another  E38 MeV particle with a mass of about 38 MeV
observed in the $\gamma \gamma$ invariant mass spectrum in
high-energy $p$C, $d$C, $d$Cu collisions at Dubna \cite{Abr12,Abr19}.
Again, the matching of the predicted mass of the isovector QED meson
with the mass of the hypothetical E38 particle, within the
experimental and theoretical uncertainties, makes the isovector
1(0$^-)$ QED meson a good candidate for the E38 particle.  There are furthermore 
possible
structures in the $\gamma \gamma$ invariant mass  spectra  in the raw
data at 10\,$-$\,15 MeV and 38 MeV in high-energy $pp$$\to$$p (\pi^+\pi^-
\gamma\gamma) p$ and $\pi^-p$$\to$$\pi^- (\pi^+\pi^- \gamma\gamma) p$
reactions at {\rm COMPASS}
\cite{Ber11,Ber14,Sch11,Sch12,Ber12,Bev12,Bev20}.   As the regions of
10\,$-$\,15 MeV and 38 MeV boson masses coincide with the predicted masses
of the isoscalar and isovector QED mesons,   a careful investigation of  
this region of low $\gamma \gamma$ invariant masses will be of great interest.

Whether the isoscalar QED meson is related to the $I(J^\pi)$=0($1^+)$
X17 particle observed in the decay of $^8$Be remains to be
investigated. If the $0(1^+)$ excited state of $^8$Be emits the
particle in the $l$=1 partial wave state, then the emitted particle
can be a $0(0^-)$ particle, the same as the X17 particle observed in
the $^4$He decay.  It will be necessary to check experimentally how
the $^8$Be nucleus in the excited $0(1^{+})$ state decays.  Additional
nuclear experiments to confirm $^8$Be and $^4$He measurements in
\cite{Kra16,Kra19} will also be of interest.

The possible occurrence of the X17 particle and QED mesons receives indirect
support from the presence of $e^+e^-$ excesses observed in the
anomalous soft photon phenomenon.  In particular, the structure of the
anomalous soft photon $dN /dp_T$ spectrum in the region of 12-20 MeV/c
 in high-energy $pp$ collisions at $p_{\rm lab}$= 450
GeV/c \cite{Bel02}   
is consistent with a possible production of a particle with a mass
around 17 MeV.
The structure of the $dN /dp_T$ spectrum in the region of 40 MeV/c
in  $e^+e^-$ annihilation at the $Z_0$
  mass of 91.18 GeV from 
the DELPHI Collaboration \cite{DEL06}
is consistent with a possible production of a particle with a mass
around 38 MeV.
  The studies of the X17 and E38 particles and
the anomalous soft photons appear to  be intimately connected.
Other similar searches can be carried out in experiments where hadrons
are produced, in high-energy hadron-$p$, $pp$, $pA$, and $AA$
collisions as well as high-energy $e^+e^-$ annihilations.  The
indirect support suggests further needs to study the invariant masses
of $e^+e^-$ and $\gamma\gamma$ pairs  in the low $p_T$
regions.

In our first exploration of the QED mesons, we have limited our
studies only to the $S$=0, $L$=0 and $I_3$=0 states.  Because of the
composite nature of such QED mesons, other collective
 rotaional and vibrational states of the open string
 with different $L$, $S$ and other quantum numbers are also possible.  It will be of great interest to extend
the frontier of QED mesons into new regions  both
theoretically and experimentally.
 
We would like to address the relevance of the QED mesons with regard
to the production of dark matter.  We envisage that in the early
evolution of the universe after the big bang, the universe will go
through the stage of quark-gluon plasma production with deconfined
quarks and gluons.  As the primordial matter cools down the
quark-gluon plasma undergoes a phase transition from the deconfined
phase to the confined phase, and hadronization occurs.  Hadrons are
then produced from the quark-gluon plasma by way of flux tube
production and string fragmentation.  As quarks and antiquarks
interact with both QCD and QED interactions, the hadron production of
QCD mesons will be accompanied by QED meson production, just as the
production of hadrons is accompanied by the production of anomalous
soft photons in high-energy hadron-nucleon collisions and $e^+$$e^-$
annihilations observed experimentally
\cite{Chl84,Bot91,Bot91,Ban93,Bel97,Bel02,DEL06,DEL08,Per09,DEL10}.
The produced QED mesons are presumably tightly bound and
non-interacting in the Schwinger's picture of two-dimensional
space-time.  In the physical four dimensional space-time with
perturbative residual interactions, they can however decay into
photons or $e^+e^-$ pairs.  Thus, gravitating assemblies of QED mesons
are $e^+e^-$ emitters and gamma-ray emitters.  On the other hand, if
they find themselves in spatial locations where their gravitational
binding energies exceed their rest masses, then their decay into
photons or electron-positron pairs will be inhibited.  There can be
QED meson assemblies produced at this stage where the gravitational
binding energies of the QED mesons exceed their rest masses.  For such
QED meson assemblies, the QED mesons will be stable against particle
decays and photon emissions.  They may form a part of the primordial cold dark
matter that may be the source of gravitational attraction for other
objects.

The recent inclusive experiment of the NA64 Collaboration \cite{Ban18}
of finding no ``dark" soft photon excesses needs to reconcile with the
earlier finding of anomalous soft photon excesses in high energy
hadron-$p$, $pp$, and $e^+$$e^-$ collisions
\cite{Chl84,Bot91,Ban93,Bel97,Bel02,DEL06,DEL08,Per09,DEL10}.  In the
detection of anomalous soft photons in the DELPHI Collaboration
\cite{DEL10}, the photon detection is carried out by studying the
electron and positron tracks in a TPC, while in the NA64 experiment,
the produced soft photon needs to penetrate a calorimeter and is
detected downstream in a separate calorimeter.  It is not known
whether the difference in the detection setups and techniques may
account for the presence or absence of excess soft photons in the two
measurements.

Future work also calls for experimental and theoretical studies of the
properties of the X17 and E38 particles and their reactions.  Much
theoretical work will need to be done to study the decays, the
properties, and the reactions of QED mesons in free space and in strong
gravitational fields to shed more lights on the fate of the QED meson
assembly in the possible primordial dark matter environment.

As it is suggested here that the hadronization at the early history of
the universe in the quark-gluon plasma phase generates simultaneously
the QED meson assemblies as seeds for primordial dark matter, it will
be of great interest to study whether QED mesons as excess $e^+e^-$
and $\gamma \gamma$ 
pairs with various invariant masses are produced in high-energy
heavy-ion collisions where quark gluon plasma may be produced.

\section{Acknowledgments}

The author wishes to thank Prof. Y. Jack Ng for helpful communications
and encouragement.  The author would like to thank
Profs. A. Koshelkin, X. Artru, I. Y. Lee, Gang Wang, Xiguang Cao,
R. Varner, K. F. Liu, V. Perelpelitsa, S. Sorensen, W. R. Hix, and
J. C. Peng for helpful discussions.  The research was supported in
part by the Division of Nuclear Physics, U.S. Department of Energy
under Contract DE-AC05-00OR22725.

\end{document}